\documentclass[
    oneside,
    aps,
    pra,
    superscriptaddress,
    twocolumn,
    a4paper,
    floatfix,
    longbibliography,
    nofootinbib
]{revtex4-2}

\usepackage[american]{babel}
\usepackage[utf8x]{inputenc}
\usepackage[T1]{fontenc}

\usepackage{grffile} % allow for period characters in filename

\usepackage{newtxtext,newtxmath}
\usepackage{microtype}

\usepackage{amsmath}
\usepackage{mathrsfs}

\usepackage{bbm}
\usepackage{bm}
\usepackage{mathtools}
\usepackage{dsfont}
\usepackage{braket}
\usepackage{cancel}
\usepackage{slashed}
\usepackage{upgreek}

\usepackage{graphicx}
\usepackage[svgnames]{xcolor}

\usepackage{siunitx}
\DeclareSIUnit\Gal{Gal}

\usepackage{makecell} %for linebreaks in latex equations in tables
\usepackage{booktabs}

\usepackage{ragged2e}

\usepackage{ragged2e}
\usepackage{array}
\usepackage{tabularx}
\usepackage{booktabs}

\definecolor{cset-aps-blueberry}{RGB}{28,128,158}
\definecolor{cset-aps-blue}{RGB}{46,44,184}
\definecolor{cset-aps-turquoise}{RGB}{0,67,88}
\definecolor{cset-aps-limegreen}{RGB}{190,219,67}
\definecolor{cset-aps-green}{RGB}{31,138,112}
\definecolor{cset-aps-yellow}{RGB}{255,225,25}
\definecolor{cset-aps-orange}{RGB}{253,116,0}
\definecolor{cset-aps-red}{RGB}{219,0,43}

\definecolor{blau}{HTML}{1575B9}
\definecolor{hellblau}{HTML}{65B7EF}
\definecolor{rot}{HTML}{E31B0A}
\definecolor{hellrot}{HTML}{FC6761}
\definecolor{gruen}{HTML}{25A131}
\definecolor{lila}{HTML}{2E2CB8}
\definecolor{grau}{HTML}{A6A6A6}

\usepackage{pgfplots}
\usepackage{tikz}
\pgfplotsset{compat=1.18}
\usetikzlibrary{math}
\pgfplotsset{
    every axis legend/.append style={
        cells={anchor=west},
        at={(0.96,0.04)},
        anchor=south east,
        font=\scriptsize,
        },
    every axis/.append style={
        },
    xmajorgrids=true,
    xminorgrids=false,
    minor x tick num=1,
    /pgf/declare function={
        shotnoise(\nO) = (1 + 5*\nO + 6*\nO^2 + sqrt(13 + 43*\nO + 48*\nO^2 + 16*\nO^3))/(2*(4 + 11*\nO + 9*\nO^2));
    },
}
\usetikzlibrary{decorations}
\usetikzlibrary{calc}

\usepackage{pict2e,picture}

\makeatletter
\DeclareRobustCommand{\Arrow}[1][]{
    \check@mathfonts
    \if\relax\detokenize{#1}\relax
        \settowidth{\dimen@}{$\m@th\rightarrow$}
    \else
        \setlength{\dimen@}{#1}
    \fi
    \sbox\z@{\usefont{U}{lasy}{m}{n}\symbol{41}}
    \begin{picture}(\dimen@,\ht\z@)
        \roundcap
        \put(\dimexpr\dimen@-.7\wd\z@,0){\usebox\z@}
        \put(0,\fontdimen22\textfont2){\line(1,0){\dimen@}}
    \end{picture}
}
\makeatother

\usepackage{hyperref}
\hypersetup{
    colorlinks=true,
    linkcolor={cset-aps-red},
    linkbordercolor={cset-aps-red},
    filecolor={cset-aps-orange},
    filebordercolor={cset-aps-orange},
    citecolor={cset-aps-blue},
    citebordercolor={cset-aps-blue},
    urlcolor={cset-aps-green},
    urlbordercolor={cset-aps-green},
    menucolor={cset-aps-limegreen},
    menubordercolor={cset-aps-limegreen},
    breaklinks=true,
    pdfborderstyle={/S/U/W 2},
    pdfpagemode=UseOutlines,
    pdfstartpage={1},
}

\DeclareMathOperator{\diag}{diag}

\newcommand{\ee}{\text{e}}
\newcommand{\ii}{\text{i}}
\newcommand{\dd}{\text{d}}

\newcommand{\PL}{\Phi_\text L}
\newcommand{\PC}{\Phi_\text{A}}
\newcommand{\hPL}{\hat\Phi_\text L}

\newcommand{\dPL}{\dot\Phi_\text L}
\newcommand{\dPC}{\dot\Phi_\text{A}}
\renewcommand{\vec}[1]{\boldsymbol{#1}}

\newcommand{\ie}{i.\,e.,}

\newcommand{\wrt}{w.\,r.\,t.}

\newcommand{\vp}{p}

\newcommand{\mapto}{\rightarrow}
\newcommand{\id}{\hat{\mathds{1}}}

\usepackage{wasysym}

\usepackage[clock]{ifsym}
\usepackage{lipsum}

\usepackage{layouts}

\newcommand{\orcid}[1]{\href{https://orcid.org/#1}{\includegraphics[width=7pt]{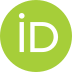}}}

\newcommand{\affTUDa}{
    \href{https://ror.org/05n911h24}{Technische Universit{\"a}t Darmstadt},
    Fachbereich Physik,
    Institut f{\"u}r Angewandte Physik,
    Schlo{\ss}gartenstr. 7,
    D-64289 Darmstadt,
    Germany}

\begin{document}

\title{
Finite-Speed-of-Light Effects in Atom Interferometry:\\
Diffraction Mechanisms and Resonance Conditions
}

\author{Christian Niehof\,\orcid{0009-0007-9628-5088}\,}
\email{cn.niehof@gmail.com}
\affiliation{\affTUDa}

\author{Daniel Derr\,\orcid{0000-0002-8690-3897}\,}
\affiliation{\affTUDa}

\author{Enno Giese\,\orcid{0000-0002-1126-6352}\,}
\affiliation{\affTUDa}

\begin{abstract}
Light-pulse atom interferometers serve as tools for high-precision metrology and are targeting measurements of relativistic effects.
This development is facilitated by extended interrogation times and large-momentum-transfer techniques generating quantum superpositions of both interferometer arms on large distances. 
Due to the finite speed of light, diffracting light pulses cannot interact simultaneously with both arms, inducing phase perturbations that compromise the accuracy of the sensor -- an effect that becomes progressively important as spatial separations increase.
For a consistent framework, we develop a theory for finite-speed-of-light effects in atom interferometers alongside with other relativistic effects such as the mass defect.
Our analysis shows that their magnitude depends crucially on the diffraction mechanism and the specific interferometer geometry.
We demonstrate that the velocity of the atomic cloud at the mirror pulse of a Mach-Zehnder interferometer is less critical than the precise tuning of the lasers for resonant diffraction.
Finally, we propose an experiment to test our predictions based on recoilless transitions and discuss mitigation strategies to reduce the bias in gravimetric applications.
\end{abstract}

\maketitle

\section{Introduction}
Due to the inherent delocalization of light-pulse atom interferometers as quantum sensors~\cite{Bongs2019} with arm separations exceeding half a meter~\cite{Kovachy2015, Abe2021}, the finite speed of light (FSL) prevents the simultaneous diffraction of both interferometer arms~\cite{Debavelaere2024}.
Since this effect induces contributions to the interferometer phase, it must be included in the discussion of relativistic corrections~\cite{Dimopoulos2008, Liu2024}.
By developing a formalism valid for any interferometer geometry, we demonstrate in this article that the phase induced by FSL critically depends on the mechanism used to diffract the atoms.
In particular, we show that the offset caused by relativistic effects in gravimetric applications is suppressed for resonantly diffracted atoms.

Atom interferometers based on single-photon transitions (SPTs)~\cite{Hu2017, Hu2019, Rudolph2020} have caught attention for potential applications in terrestrial large-baseline setups~\cite{Abend2024,Abdalla2025,DiPumpo2024} for gravitational-wave~\cite{Graham2013,Badurina2025} and dark-matter detectors~\cite{Arvanitaki2018,Derr2023}, as well as in other practical scenarios beyond fundamental physics~\cite{Bongs2019}.
In ground-based experiments, the laser frequency must be chirped to remain on resonance with the atomic transition~\cite{Marzlin1996,Bott2023,Boehringer2024}.
The chirp leads to different transferred momenta and thus impacts the detected interferometric phase.
The mass defect, \ie{} different masses for the ground and excited state, leads to an additional relativistic perturbation of the phase that becomes even more important when working with optical transitions.
Consequently, a theoretical framework incorporating FSL effects requires also the inclusion of other relativistic perturbations, as they may contribute at the same order of magnitude~\cite{Dimopoulos2008, Tan2017relativistic, Liu2024}.
Moreover, different types of atom optics and atomic transitions are crucial to isolate terms like the gravitational redshift~\cite{DiPumpo2021, Ufrecht2020ugr, Roura2020} and other contributions in tests of general relativity~\cite{DiPumpo2023} or quantum clock interferometry~\cite{Loriani2019, Roura2021, Pikovski2017, Zych2011}.

Also in two-photon transitions (TPTs) used for Raman or Bragg diffraction~\cite{Hartmann2020}, frequency chirps are necessary in vertical setups to ensure resonant transitions.
This frequency chirp is in general associated with a modified momentum transfer, which has been experimentally observed~\cite{Peters2001, Cheng2015}.
However, in contrast to SPTs such effects can be suppressed by chirping counterpropagating light fields in opposite directions.
But even in this case, the mean velocity at the time of the mirror pulse of a Mach-Zehnder interferometer (MZI) has been identified as the dominant effect~\cite{Tan2016, Tan2017}, an insight that was predicted to impact tests of the equivalence principle~\cite{Xu2022}.

In this article, we consider FSL, chirped lasers, and the mass defect~\cite{Asano2024,Sonnleitner2018,Schwartz2019,Yudin2018,Martinez2022}, which implies time dilation and gravitational redshift.
We demonstrate that the phase depends crucially on the diffraction mechanism and the greatest impact occurs in SPT configurations.
By allowing for a deviation from the resonance condition, we show that relativistic effects are suppressed on resonance and that the mean velocity at the time of the mirror pulse does not contribute to the offset in gravity measurement.
Previous results~\cite{Dimopoulos2008, Tan2016, Tan2017, Tan2017relativistic, Liu2024} emerge naturally as limiting cases of our framework.
Furthermore, we introduce a timing asymmetry~\cite{Tan2017} at the mirror pulse~\cite{Muentinga2013} as a mitigation strategy in analogy to rotation~\cite{Dickerson2013, Lan2012} and gravity-gradient~\cite{Roura2017,Overstreet2018,DAmico2017,Ufrecht2021GG} compensation methods to suppress a dependence on initial conditions.

We begin by deriving the relativistic propagation~\cite{DiPumpo2022} of chirped light pulses in weak gravity in Sec.~\ref{sec:LightGrav}.
These insights are used in Sec.~\ref{sec:Ufrecht} to calculate the perturbations of the interferometer phase induced by chirped laser pulses and relativistic effects in gravity.
The result is then applied to an MZI used as gravimeter in Sec.~\ref{sec:MZI} and includes a discussion of different diffraction mechanism like SPTs, as well as TPTs in the form of Bragg and Raman diffraction.
In Sec.~\ref{sec:beyondMZI}, we study FSL effects in a butterfly geometry and propose an FSL test based on recoilless transitions~\cite{Alden2014, Janson2024} in Sec.~\ref{sec:E1M1}.
The article concludes with a discussion about orders of magnitudes of the phase offset in Sec.~\ref{sec:conc}.
The mitigation scheme for initial conditions is introduced in Appendix~\ref{apx:comp}.

\section{Chirped light pulses in gravity}
\label{sec:LightGrav}
When performing atom-interferometric experiments in a terrestrial laboratory, not only the atoms are subject to gravity, but so are the light pulses used for atom-optical manipulation.
Here, we study the influence of a weakly curved spacetime in the form of the Rindler metric in the lab frame $(g_{\mu\nu})=\diag\bigl((1+gz/c^2)^2,-1,-1,-1\bigr)$ encoding effects of the gravitational acceleration $g$ on the phase of a laser beam, where $c$ denotes the speed of light, and $z$ the height.
In this case, the phase $\Phi$ of the beam follows from the eikonal equation~\cite{DiPumpo2022,Dolan2018}
\begin{equation}
    0
    =g^{\mu\nu} (\partial_\mu\Phi) (\partial_\nu\Phi)
    =(1+gz /c^2)^{-2}(\partial_0\Phi)^2-(\vec{\nabla}\Phi)^2.
    \label{eq:Eikonal}
\end{equation}
This partial differential equation has two solutions $\Phi_\pm$ corresponding to upwards ($+$) and downwards ($-$) propagating light beams with different initial conditions.
We assume that all lasers are aligned along the $z$ axis determined by gravity.
These two solutions can be used to describe different situations: 
(i) SPTs induced by the interaction of the atoms with a single laser and (ii) TPTs induced by the interaction with two counterpropagating lasers of different frequencies.

\begin{figure}[ht]
    \centering
    \includegraphics[page=1]{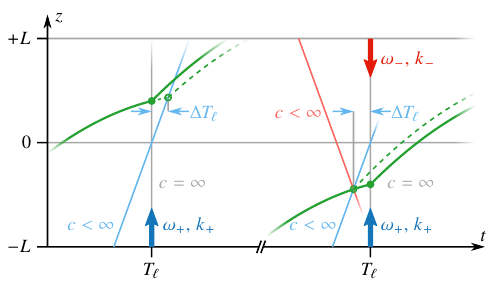}
    \caption{
        Spacetime diagram showing the impact of FSL effects in SPTs (left) and TPTs (right).
        Neglecting FSL effects (gray, $c=\infty$), both schemes diffract the atoms instantaneously at $T_\ell$, changing the trajectory (solid green).
        The upwards pointing laser (frequency $\omega_+$, wave vector $k_+$, blue, positioned at $z=-L$) is pulsed and controls the time of interaction \wrt{} the reference height $z=0$.
        The light cones accounting for FSL (light blue, $c<\infty$) indicate that the switching of the lasers must be advanced to $T_\ell$.
        Furthermore, the time of interaction shifts by $\Delta T_\ell$, resulting in perturbed atomic trajectories (dashed green).
        For TPTs, a second laser (frequency $\omega_-$, wave vector $k_-$, red, positioned at $z=+L$) is pointed downwards in addition.
        }
    \label{fig:SPTvsTPT}
\end{figure}

For SPTs, shown in the spacetime diagram of Fig.~\ref{fig:SPTvsTPT} on the left, a single, upwards-pointing laser of frequency  $\omega_+$ displayed in blue and positioned at $z=-L$ interacts with moving atoms by imprinting its phase and transferring its momentum $\hbar k_+$, where $k_+$ is the laser's wave vector.
When neglecting FSL ($c=\infty$, gray), this momentum transfer of the $\ell$th light pulse at time $T_\ell$ modifies the idealized atomic trajectory (solid green).
However, the exact time of this interaction including FSL ($c<\infty$, blue light cone) is determined by the intersection of the light cone with the atomic trajectory and leads to a position-dependent time delay $\Delta T_\ell$.
We assume that the laser switching is timed such that the pulse passes $z=0$ at time $T_\ell$.
The delay $\Delta T_\ell$ will induce an atomic trajectory (green dashed) slightly perturbed from the idealized trajectory.

In the situation displayed on the right of Fig.~\ref{fig:SPTvsTPT}, an additional second downwards-pointing laser (red) with frequency $\omega_-$ and wave vector $k_-$, positioned at $z=L$ is used to drive TPTs and transfer the total momentum $\hbar K=\hbar(k_++k_-)$.
Since the time of interaction is determined by the blue control laser, the passive red laser does not influence the time of interaction, even though both lasers might be switched on at different times~\cite{Tan2016, Dimopoulos2008, Wang2021}.

For optimal transitions, the frequencies of the light fields have to be matched to the transferred energy.
Because atoms are subject to gravity and they are therefore accelerated in the lab frame, their time-dependent Doppler shift~\cite{Marzlin1996} has to be compensated by the light fields to remain on resonance.
Thus, their frequencies $\omega_\pm$ are chirped by rates $\omega_\pm\sigma_\pm/c$, where $\sigma_\pm$ are given in units of acceleration.
For the light phase, we make therefore the ansatz
\begin{equation}
\label{eq:chirped_phase_in_grav}
    \Phi_\pm(z,t)=\phi_\pm+\kappa_\pm(z,t)- \omega_\pm (t-t_i) \left[1+\frac{\sigma_\pm(t-t_i) }{2c}\right],
\end{equation}
which we separate into a spatial $\kappa_\pm(z,t)$ and purely time-dependent part.
We use the initial conditions $\kappa_\pm(\mp L, t)=0$ such that the pulses are initiated at positions $z=\mp L$ at time $t=t_i$ with initial phases $\phi_\pm=\Phi_\pm(0,t_i)$ in agreement with Fig.~\ref{fig:SPTvsTPT}.
Equation~\eqref{eq:Eikonal} is solved by the spatial part
\begin{equation}
    \kappa_\pm= \pm (z\pm L) k_\pm \Big[ 1 + \frac{\sigma_\pm (t-t_i)}{c} - \frac{(g\pm \sigma_\pm)z + (\sigma_\pm \mp g ) L}{2c^2} \Big]
\end{equation}
up to order $\mathcal O(c^{-2})$.
Since the frequency chirp implies also a change of the associated wave vector, we observe a modification that increases linearly in time.
In addition, chirping causes an accelerational redshift~\cite{Okolow2020} of the light phase scaling quadratically with position, which adds to the gravitational redshift caused by general-relativistic light propagation in the Rindler metric.

Atomic diffraction relying on TPTs like Raman and Bragg diffraction is induced by subsequent absorption and emission of photons from both counterpropagating light beams.
These transitions lead to an effective light phase
\begin{equation}
    \Phi_\text{eff}(z,t)=\Phi_+(z,t)-\Phi_-(z,t)= \PL(z,t)+\delta\Phi(z,t),\label{eq:Phi_eff}
\end{equation}
which can be divided into a dominant light phase $\Phi_\text L$ and relativistic corrections $\delta\Phi$.
Note that the limit of SPTs can be recovered by setting $\Phi_-=0$, while recoilless E1-M1 TPTs that rely on the absorption of photons from both light beams are in principle also included by choosing opposite signs for $\omega_-$, $k_-$ and $\phi_-$ compared to the laser traveling upwards.
The dominant light phase is given by
\begin{equation}
     \PL(z,t)= \Phi_\text{off}+ K z- \Delta \omega (t- t_i^\prime) + K \sigma (t- t_i^\prime)^2/2
\end{equation}
with initial phase $\Phi_\text{off}=\Phi_\text{eff}(0, t_i^\prime)$, effective two-photon wave vector $K=k_+ + k_-$ corresponding to the transferred momentum, frequency difference $\Delta \omega = \omega_+-\omega_-=c\Delta k$ corresponding to the transferred energy, and modified initial time $t_i^\prime=t_i+L/c$.
Thus, all laser pulses are giving a lead in time by $L/c$ to account for the propagation from the laser sources at $z=\pm L$ to the center of the experimental setup at $z=0$, as already anticipated in Fig.~\ref{fig:SPTvsTPT}.
While it is possible to keep two different chirping parameters~\cite{Tan2016, Tan2017, Tan2017relativistic} we have chosen opposite chirping, \ie{} $\sigma_\pm=\mp\sigma$, which has been shown to suppress FSL effects~\cite{Peters2001, Cheng2015}.
The last contribution of $\PL$ is used to lock the frequency chirp $\sigma$ to gravity in gravimetric applications, see Sec.~\ref{sec:MZI}.

Following the solutions of the individual phases, the effective phase experiences the perturbation
\begin{equation}
\label{eq:delta_Phi}
     \delta \Phi(z,t)= -\Delta kz\frac{\sigma(t-t_i^\prime )}{c}-Kz\frac{(g-\sigma)z}{2c^2}+\ldots
\end{equation}
that includes a chirp modification of the transferred momentum, which scales with the difference of the wave vectors $\Delta k= k_+-k_-=\Delta \omega /c$.
If chirping is not performed perfectly in opposite directions, the resulting wave-vector chirp may become more pronounced, a phenomenon that has already been experimentally observed~\cite{Peters2001, Cheng2015}.
In addition, the combination of gravitational and accelerational redshift of the laser phase that contributes to the phase perturbation scales quadratically in position.

In general, $K$ and $\Delta \omega$ are two independent parameters for TPTs and are not connected by a dispersion relation.
However, we find $K=\Delta k=\Delta\omega/c$ for SPTs with $k_-=0=\omega_-$.
As a consequence, the first term of $\delta \Phi$ may contribute significantly for SPTs, while the second term is suppressed for perfect chirping ($\sigma=g$) or for recoilless transitions with $k_-=-k_+$, i.\,e., $K=0$.
Table~\ref{tab:Disp} summarizes the parameters, characteristics, and resonance conditions of single-photon, Bragg, Raman, and E1-M1 recoilless transitions.

\begin{table}[ht]
    \centering
    \caption{
    Resonance condition, magnitude and identities of the effective wave vector $K$, giving rise to the recoil, as well as the internal atomic energy $\hbar \omega_\text{A}$, are listed for different diffraction mechanisms.
    Here, we used $\Delta \omega = c \Delta k$, $v_\text{R}$ denotes the resonant velocity and $v_K=\hbar K /\bar m$ is the recoil velocity for an atom of mean mass $\bar m$.
    }
    \begin{tabular}{llll}
    \hline\hline
        Mechanism & Resonance condition& Recoil & Int.\,energy\\
    \hline
        SPT & 
            $\Delta k=k_\text A+K(v_\text R+v_\text K/2)/c$ &
            $K=\Delta k$ &
            $\omega_\text A\cong\Delta\omega$\\
        TPT: Bragg &
            $\Delta k=K(v_\text R+v_\text K/2)/c$ &
            $K\ggg\Delta k$ &
            $\omega_\text A\equiv0$\\
        TPT: Raman &
            $\Delta k=k_\text A+K(v_\text R+v_\text K/2)/c$ &
            $K\gg\Delta k$ &
            $\omega_\text A\ll Kc$\\
        TPT: E1-M1 &
            $\Delta k=k_\text A$ &
            $K=0$ &
            $\omega_\text A=\Delta\omega$\\
    \hline\hline
    \end{tabular}
    \label{tab:Disp}
\end{table}

\begin{figure}[ht]
    \centering
    \includegraphics[page=2]{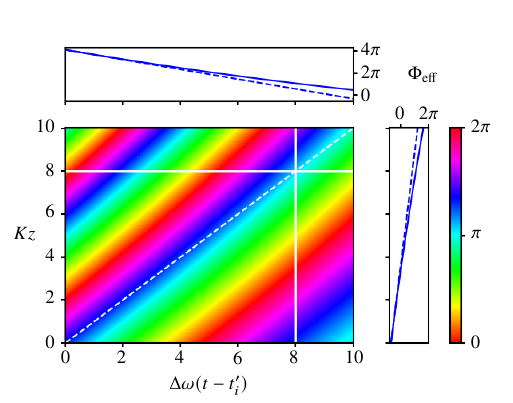}
    \caption{
        The effective light phase $\Phi_\text{eff}$ from Eq.~\eqref{eq:Phi_eff} is shown for an upwards propagating laser ($\Phi_-\equiv0$) in a dimensionless spacetime diagram with $\Phi_\text{off}=4\pi/3$, $\sigma/(c\Delta\omega)=0.05$, and $g/\sigma=2$.
        The phases along the solid white lines are plotted as solid blue lines in the panels aside and show a nonlinear deviation from the linear phase contributions (dashed blue) due to redshift (right panel) and chirping (top panel).
        The dashed white line represents the light cone in flat Minkowski space, guiding the viewer's eye.
        }
    \label{fig:LightProp}
\end{figure}

The effective phase $\Phi_\text{eff}$ is plotted in Fig.~\ref{fig:LightProp} in a dimensionless spacetime diagram for a single upwards-pointing laser.
The line of constant phase deviates from the light cone in flat Minkowski spacetime (dashed white line).
The top and side panels are cuts along the white lines and show the nonlinearity introduced by chirping and gravity in comparison to the dominant linear terms (dashed blue lines).

To gain more insight in $\Phi_\text{eff}$ and to find a connection between the resonance condition, initial conditions, and chirping, we consider a two-level atom with energy splitting $\hbar\omega_\text A$ moving along the trajectory $z(t)=z_0+v_0(t-t_i^\prime)-g(t-t_i^\prime)^2/2$ with initial position $z_0$ and initial velocity $v_0$.
The coordinate system is defined such that neither redshift nor chirp is present at the initial time $t=t_i^\prime$ and position $z=0$.
Inserting this trajectory into $\PL$, the gravitational acceleration $g$ of the atom  included in $Kz(t)$ cancels the chirp for $\sigma=g$.
As already pointed out, the accelerational redshift~\cite{Okolow2020} cancels the gravitational redshift $\delta\Phi$ in this case as well.
Moreover, the initial velocity $v_0$ leads to a phase contribution $K v_0 (t-t_i^\prime)$, that corresponds to a static Doppler shift of frequency $ K v_0 $.
In fact, the explicit time dependence of the trajectory modifies the transferred energy, which is given by $-\hbar\partial_t\Phi_\text{eff}(z(t),t)$, while the transferred momentum is given by $\hbar \partial_z\Phi_\text{eff}(z(t),t)$.
Since the energy $\hbar \Delta \omega$ of the light has to account for the internal energy $\hbar \omega_\text{A}$ and the transferred kinetic energy during diffraction, the resonance condition up to  $\mathcal O(c^{-1})$ for an atom with initial velocity  $v_\text{R}$ takes the form
\begin{equation}
\label{eq:res_cond}
    \Delta \omega = \omega_\text{A} + \omega_K + \omega_\text{D}(v_\text{R}),
\end{equation}
where we introduced the Doppler frequency $\omega_\text{D}(v_\text{R})=Kv_\text{R} $, the recoil frequency $\omega_K= K v_K/2$ and the recoil velocity $v_K = \hbar K/\bar m$, which accounts for the kinetic energy gained by the recoil during diffraction.
Note that we introduced a resonant velocity $v_\text R$ that fulfills Eq.~\eqref{eq:res_cond}, as any deviation from this velocity leads to a Doppler detuning and the initial velocity $v_0$ of the atom might not be perfectly known.
In addition, a Doppler detuning over the velocity distribution of the atom leads to velocity selectivity~\cite{Giese2015,Boehringer2024}.
Hence, an atom is only diffracted resonantly at time $t=t_i^\prime$ and position $z=0$ for $v_0=v_\text R$.
Since the acceleration of the atom can be compensated by chirping, the lasers remain resonant during the whole interferometer sequence.
The resonant velocity is not needed in the case of recoilless transitions ($K=0$), as they are Doppler free ($\omega_\text D(v_\text{R})=0$ for all $v_\text{R}$).

The first contribution to $\delta\Phi$ includes the wave-vector modification induced by chirping and is proportional to $\Delta k = \Delta \omega/c$ and can be connected by the resonance condition from Eq.~\eqref{eq:res_cond} to the atomic wave vector $k_\text A=\omega_\text A/c$ and Doppler as well as recoil terms.
As summarized in Table~\ref{tab:Disp}, the resonance condition implies that $\delta\Phi$ is small for Raman and Bragg transitions, because either the atomic wave vector vanishes for state-preserving Bragg transitions ($k_\text A\equiv0$), or it is small for Raman transitions between two hyperfine states, typically separated by gigahertz.
However, in the case of optical SPT $\delta \Phi$ becomes significant as $\Delta k$ takes values of hundreds of terahertz.

To generate an interferometer with two arms, the atoms are diffracted at specific times controlled by a sequence of light pulses.
Since the times of this interaction on each arm are necessary for the description the whole sequence, we consider an SPT with an upwards-propagating light pulse modeled by a Gaussian wave packed of spectral width $\sigma_\omega$.
The interaction time is obtained from Fourier-transforming the Gaussian spectrum by multiplying it with the light phase $\exp(\ii\Phi_+)$ as a solution of Eq.~\eqref{eq:Eikonal} and integrating over all frequencies $\omega_+$.
Since all frequency components are chirped in the same way, the chirping parameter $\omega_+\sigma_+/(2c)$ is kept constant, as $\omega_+/c$ was only introduced artificially to write the chirping parameter $\sigma$ in units of an acceleration.
The positive-frequency part of the resulting electric field is given by
\begin{align}
    \vec E(z,t)
    =\vec E_0\exp\bigl\{
        -\sigma_\omega^2(t-t_i'-z/c)^2/2
    \bigr\}\ee^{
        \ii\Phi_+(z,t)
    }
    \label{eq:EnvelopeLight}
\end{align}
including corrections up to order $\mathcal O(c^{-2})$, where now the frequency $\omega_+$ denotes the central frequency of the pulse  initiated at $t=t_i^\prime$.
This expression describes the propagation of such a pulse in space and time. 

The dipole interaction of the light pulse with the atom is then obtained within the rotating-wave approximation from the operator-valued and time-dependent Rabi frequency $\Omega_\text R (\hat z, t) =-\vec d\cdot\vec E(\hat z, t)/(2\hbar)$, with the atomic dipole moment $\vec d$.
Within the dipole approximation, the electric field, and by that the Rabi frequency, depends on the center-of-mass position operator $\hat z$ of the atom.
In the Delta-pulse approximation~\cite{Schleich2013} $\sigma_\omega\mapto0$, the envelope from Eq.~\eqref{eq:EnvelopeLight} is proportional to a Dirac Delta at $t=t_i^\prime+\hat z/c$.
Thus, for a pulse initiated at time $t_i^\prime=T_\ell$, interaction takes place at $t=T_\ell+\hat z/c$, which depends on the operator-valued position of the atom.

For a TPT, we take a passive background laser~\cite{Tan2016, Dimopoulos2008, Wang2021} with phase $\Phi_-$ and a pulsed control laser with phase $\Phi_+$.
Since the two-photon Rabi frequency~\cite{Giese2015} scales with $(\vec d \cdot\vec E_+)(\vec d \cdot \vec E_-)$, the phase of the light field has to be replaced by $\Phi_\text{eff}$, leading to the same expression for the time delay.

Upon the interaction, atoms are diffracted and the momentum $\hbar K$ is transferred, which is apparent from the displacement operator $\exp(\ii K \hat z)$ included in the dominant part of the effective light phase $\Phi_\text{eff}$.
But also the remainder of $\Phi_\text{eff}$ is imprinted on the diffracted atoms.
The shift $\hat z/c$ in the envelope accounts for FSL, so that the time of interaction depends on the specific arm.
Assuming instantaneous transitions, this interaction can be described by an arm-dependent effective interaction potential $-\hbar\dot\lambda\Phi_\text{eff}(\hat{z},t)$.
Here, the term $\dot\lambda$ is a sequence of of Dirac Deltas centered around the individual pulses at interaction times $T_\ell+\hat z/c$.
The sign of $\dot\lambda$ accounts for the direction of momentum transfer and is associated with photon absorption $(+)$ and emission $(-)$ and the corresponding two-photon process.
As $\dot\lambda$ originates from the change of the internal state, the state-response function $\lambda$ switches between $\lambda=-1/2$ for the atom being in the ground state and $\lambda=+1/2$ for the atom being in the excited state, see Fig.~\ref{fig:Lambda}.
The state-response function can be defined in a similar way for Bragg transitions that only transfer momentum by choosing $\omega_\text{A}=0$.
The time derivative of $\lambda$ produces the sequence of Dirac Deltas that accounts for instantaneous momentum transfers and phase imprints.
For simplicity, we choose $t_i^\prime=0$ in the following.

\section{Atom-interferometric phase shifts}
\label{sec:Ufrecht}
To properly integrate FSL into the derivation of the interferometer phase to order $\mathcal O(c^{-1})$, one must simultaneously address additional relativistic contributions of similar order.
Hence, we include the mass defect~\cite{Asano2024,Sonnleitner2018,Schwartz2019,Yudin2018,Martinez2022} originating from the energy-mass equivalence, which leads to a clock phase, to time dilation caused by the atomic motion, and a to redshift caused by the height of the atom in the gravitational potential.
Since the internal state may change along a path, the mass defect is modeled by a state- and path-dependent mass $m(\hat z, t)=\bar m + \lambda(\hat z, t) \Delta m$.
In a two-level atom, the energy splitting $\hbar \omega_\text{A} = \Delta m c^2$ is connected to the mass difference $\Delta m$.
Moreover, the mean mass $\bar m$ is associated with the average mass of the atom in the ground and excited state.
In a path-dependent description~\cite{Ufrecht2020,DiPumpo2023}, the effective Hamiltonian takes the form
\begin{equation}
    \hat H
    =mc^2
    +\frac{\hat{\vp}^2}{2m}-\frac{\hat{\vp}^4}{8m^3c^2}
    +m g\hat{z} -\hbar\dot\lambda(\hat z, t) \Phi_\text{eff}(\hat{z},t) 
\end{equation}
with the atom's momentum operator $\hat p$ fulfilling the canonical commutation relation $[\hat z, \hat p]=\ii\hbar \id$.
Expanding the Hamiltonian in $\Delta m/\bar m=\hbar\omega_\text A/(\bar mc^2)$ up to order $\mathcal O(c^{-2})$ leads to
\begin{equation}
    \hat{H}
    =\frac{\hat{\vp}^2}{2\bar m}
        +\bar mg\hat z
        +\lambda \frac{\Delta m}{\bar m}\left(
            \bar mc^2
            -\frac{\hat{\vp}^2}{2\bar m}
            +\bar mg\hat z
        \right)
        -\hbar\dot\lambda \bigl(
            \hat \Phi_\text{L}
            + \delta\hat\Phi \bigr).
\end{equation}
Here, we have omitted the constant rest energy $\bar mc^2$, which does not induce a phase difference between both arms, and the correction of the relativistic kinetic energy $\hat{\vp}^4/(8\bar m^3c^2)$, whose contribution is of higher order assuming that the velocity difference between both arms is proportional to the recoil velocity $v_K= \hbar K / \bar m$.

\begin{figure}[ht]
    \centering
    \includegraphics[page=3]{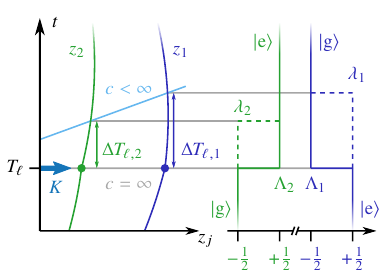}
    \caption{
        When accounting for FSL, an SPT pulse with wave vector $K$ interacting with two arms ($j=1,2$) at different positions $z_j$ and in different internal states (ground state $\ket{\text g}$, excited state $\ket{\text e}$) experiences an arm-dependent time shift $\Delta T_{\ell,j}$ of the interaction relative to the idealized case ($c = \infty$), where the interaction takes place for both arms simultaneously.
        The state-response functions $\lambda_j$ for each arm, which encode the direction of the momentum transfer and the internal state change, differ from the idealized response functions $\Lambda_j$ that neglect FSL.
        }
    \label{fig:Lambda}
\end{figure}

Next, we introduce the idealized state-response function $\Lambda$ in analogy to $\lambda$, but with the significant stipulation of infinite speed of light ($c = \infty$).
Figure~\ref{fig:Lambda} shows an example of a laser pulse with the effective wave vector $K$, initiated at time $t=T_\ell$, interacting with both arms $j=1,2$ of an atom interferometer where the atoms are in different internal states and at positions $z_j$.
Accounting for FSL ($c<\infty$), the laser pulse interacts with both interferometer arms at different times yielding two different time shifts $\Delta T_{\ell,j}$ and thus the state-response functions $\lambda_j$ associated with both arms switch values at different times.
In the idealized case ($c=\infty$), the laser pulse interacts at the same time with both interferometer arms and thus the idealized state-response functions $\Lambda_j$ change values at the same time $T_\ell$, \ie{} $\Lambda_j = \lim_{c \rightarrow \infty} \lambda_j$.

Introducing the atomic phase $\PC=\omega_\text{A} t$ that is induced by the energy splitting and assuming that the FSL time delay $\hat z/c$ can be treated as perturbation, the Hamiltonian takes the form
\begin{equation}
    \hat{H}
        =\frac{\hat{\vp}^2}{2\bar m}
        +\bar mg\hat z
        -\hbar\dot\Lambda(\hat\Phi_\text{L}+\PC)
        + \hat V,
    \label{eq:splitH}
\end{equation}
such that the unperturbed contributions include only the idealized state-response function $\Lambda$ and no relativistic effects, while all perturbations are collected in the potential
\begin{equation}
\label{eq:V_pert}
 \frac{\hat V}\hbar = 
        \lambda \omega_\text A \left[1-\frac{\hat{\vp}^2}{2\bar m ^2 c^2}+\frac{g\hat z}{c^2}\right]
        +\dot\Lambda\PC
        -\dot\lambda\,\delta\hat\Phi
        -\left(\dot\lambda-\dot\Lambda\right)\hPL.
\end{equation}
In this form, the momentum $\hbar K$ is transferred by the position-dependent light phase $K\hat z$ included in $\hat\Phi_\text L$ at times  $T_\ell$, while at the same times the dominant contribution of the light and atom phase is imprinted on the atom, omitting any FSL effects.
In fact, Eq.~\eqref{eq:V_pert} acts as a perturbation since the difference $\dot\lambda-\dot\Lambda$ scales with the time shift $\hat z/c$, which will introduce an arm-dependent time delay $\Delta T_{\ell,j}$ as seen in Fig.~\ref{fig:Lambda}, which is small for ground-based setups.
Partial integration shows that $\lambda\omega_\text A+\dot\Lambda\PC$ is small for the same reason.
Moreover, the perturbation phase $\delta\Phi$ of the light field is treated as a small quantity, as are the time-dilation term $\hat p^2/(2\bar m^2 c^2)$ and redshift term $g\hat z/c^2$ of the atom.
Hence, the potential $\hat V$ can be treated as a perturbation.

By design, atom interferometers measure the phase difference between two arms $j=1, 2$, where each arm's dynamics is characterized by different state-response functions  $\lambda_j$ and $\Lambda_j$.
As a result, the Hamiltonian from Eq.~\eqref{eq:splitH} becomes path-dependent and all respective quantities are equipped with the index $j$.
Following the perturbative operator approach~\cite{Ufrecht2020}, the total phase of the atom interferometer is given in first order by
\begin{equation}
\label{eq:phase_difference}
    \phi_\text{AI}= \phi_\text{un} + \varphi_1 - \varphi_2
\end{equation}
where the unperturbed geometry must close in phase space~\cite{Kleinert2015,Giese2014} and the unperturbed phase $\phi_\text{un}$ is calculated by setting $\hat V_j=0$.
The perturbation phase $\varphi_j=-\int\!\dd t\,V_j(z_j(t),p_j(t),t)/\hbar$ along arm $j=1, 2$ is calculated by replacing the operators $\hat z$ and $\hat p$ in $\hat V_j$ by the classical trajectories $z_j(t)$ and $p_j(t)$ along arm $j$.
They follow from the classical analogue of the Hamiltonian assuming $\hat V_j=0$ and are a consequence of solving the respective operator equations of motions in an interaction picture \wrt{} the unperturbed Hamiltonian.

In the following, the path dependence is included in the idealized state-response functions $\Lambda_j$.
For simplicity, we omit the index $j$ and reduce the discussion to a single arm.
As mentioned in Sec.~\ref{sec:LightGrav}, the expression $\dot\lambda$ is a sequence of Dirac Deltas centered around $t-T_\ell-z(t)/c$.
Solving the equation $t - T_\ell - z(t)/c =0$ successively order by order in $c$, we find the interaction times at $t\cong T_\ell+z(T_\ell)/c[1+v(T_\ell)/c]=T_\ell+\Delta T_\ell$.
Hence, we define the time delay as $\Delta T(t) = z(t)/c[1+v(t)/c]$ with velocity $v(t) = p(t) / \bar m$.
Next, we expand $\lambda\cong \Lambda - \dot \Lambda \Delta T$ and $\dot \lambda \cong \dot \Lambda - \ddot \Lambda \Delta T$ around $\Delta T_\ell\cong 0$.
To express the perturbative phase solely by step functions or Dirac Deltas, we perform a partial integration of the terms with $\ddot \Lambda$, treating $\Delta T_\ell$ as a constant at each pulse.
Similarly, we integrate the term $\Lambda \omega_\text{A}$ by parts to arrive at
\begin{equation}
\label{eq:varphi_first_step}
    \varphi=\int\!\dd t\,\Biggl\{
    \dot\Lambda\Big[\delta\Phi+(\dPL+\dPC)\Delta T\Big]
    +  \Lambda\omega_\text A\left[\frac{v^2}{2c^2}-\frac{gz}{c^2}\right]\Biggr\}
\end{equation}
for the phase along one arm.
To find this compact form, we have omitted the boundary terms from the integration by parts, as they are identical on both arms and cancel when calculating the phase difference following Eq.~\eqref{eq:phase_difference}.
We clearly see that the frequency contributions $\dot \Phi_\text{L}$ and $\dot \Phi_\text{A}$ multiplied by $\Delta T(t)$ evaluated at the times of the pulses give rise to phase contributions.
The dominant FSL effect arises from frequencies multiplied by time shifts $z/c$.
Frequencies multiplied by time shifts $zv/c^2$ are caused by the finite velocity of atoms and are referred to as FSL Doppler effect in the subsequent analysis~\cite{Roura2025}.

With the explicit expressions for $\dot \Phi_\text{L}$, $\dot \Phi_\text{A}$, and $\delta \Phi$ and assuming that the lasers point always in the same direction, which is not the case in most large-momentum-transfer schemes~\cite{Graham2013,Rudolph2020}, the expression can be further simplified by performing another integration by parts.
When we identify $\omega_\text A=ck_\text A$, we arrive at
\begin{align}
\label{eq:second_varphi}
    \varphi=\int\!\dd t\,\biggl\{
        \dot\Lambda \left[ (k_\text{A}-\Delta k)z+(K-\Delta k)z\frac{v+\sigma t}{c} \right]
        -\Lambda k_\text{A}\frac{v^2}{2c}
        \biggr\}
\end{align}
in order $\mathcal O(c^{-1})$, where we again omit boundary terms due to cancellation when calculating the phase difference between the two arms.
The first term is indeed of order $\mathcal O(c^{-1})$, as given by the resonance condition $\Delta k-k_\text A=K(v_0+v_\text K/2)/c$
from Eq.~\eqref{eq:res_cond}.
Expressions containing velocities are evaluated symmetrically at the time of the idealized pulses $T_\ell$, \ie{} taking the average of the left- and right-sided limit.
This in accordance with a symmetrically regularized Dirac Delta~\cite{Kurasov1996} which follows from the typical stipulation $\delta(t)=\delta(-t)$.
The final integration by parts has eliminated the redshift term $\omega_\text A gz/c^2$, now included in the time dilation term with a reversed sign, which highlights the connection between kinetic and potential energy analogous to the virial theorem.
The expression in Eq.~\eqref{eq:second_varphi} is valid for all closed unperturbed geometries where the direction of the momentum transfer is not changed.
It's evaluation is straightforward due to the Dirac Deltas contained in $\dot\Lambda$.

\section{Mach-Zehnder Gravimeter}
\label{sec:MZI}
As a prime example, we apply the general phase expression from Sec.~\ref{sec:Ufrecht} to an MZI used for gravimetry.
It consists of a beam splitter pulse at $t=0$, creating a superposition of two arms, followed by a mirror pulse at $t=T$ redirecting the atomic beams, and a final beam splitter pulse applied at $t=2T$ to generate the interference.
This geometry can be described by the sequence of Dirac Deltas $\dot\Lambda_1(t)=\delta(t)-\delta(t-T)$ for arm 1 and $\dot\Lambda_2(t)=\delta(t-T)-\delta(t-2T)$ for arm 2, where the idealized state-response functions $\Lambda_j$ are defined accordingly.
We list in Table~\ref{tab:PhaseContr} the unperturbed phase $\phi_\text{un}$, the integrands of Eq.~\eqref{eq:second_varphi}, and the associated phase differences, where expectation values of the velocities at discontinuities have been evaluated symmetrically as described above.
\begin{table}[ht]
    \centering
    \caption{
    Unperturbed phase $\phi_\text{un}$, phase contributions from Eq.~\eqref{eq:second_varphi}, and the resulting phase differences are listed for an MZI.
    The mean velocity at the mirror pulse is denoted by $v_\pi=v_0-gT+v_\text K/2$ with the recoil velocity $v_\text K=\hbar K/\bar m$.
    }
    \begin{tabular}{lccc}
        \hline\hline
        Term & Integrand & Phase contribution\\
        \hline
        Unperturbed &  $\bullet$ &
            $\displaystyle- K (g - \sigma) T^2 $\\
        \hline
        FSL clock & $\displaystyle\dot\Lambda(k_\text{A}-\Delta k)z$&
           $\displaystyle(\Delta k-k_\text{A})gT^2$\\
        FSL Doppler & $\displaystyle\dot\Lambda(K-\Delta k)z v/c$& 
           $(\Delta k-K) g T^2 3v_\pi/c$\\
        Chirp & $\displaystyle\dot\Lambda(K-\Delta k)z \sigma t/c$& 
           $(K-\Delta k)\sigma T^2 (2v_\pi-gT)/c$\\
        Time dilation & $\displaystyle-\Lambda k_\text{A}v^2/(2c)$&
            $\displaystyle-k_\text{A}gT^2 v_\pi/c$\\
        \hline\hline
    \end{tabular}
    \label{tab:PhaseContr}
\end{table}

As expected for a chirped MZI, the unperturbed phase takes the form $\phi_\text{un}=-K(g-\sigma)T^2$ and is complemented by four 
relativistic perturbations:

(i)
If the time spent in each internal state differs between both arms, a clock phase consisting of this timing asymmetry multiplied by $\omega_\text{A} - \Delta \omega$ is included in the signal in complete analogy to atomic clocks.
The dominant FSL effect $z/c$ generates such an asymmetry and therefore causes an FSL clock phase.
Because the involved frequencies are typically large, this shift takes a form similar to the unperturbed contribution $\phi_\text{un}$.
In fact, we observe primarily effects  for SPTs as discussed below.
This contribution is not explicitly discussed in most studies, but it has been included in relativistic treatments of atom interferometers before~\cite{Dimopoulos2008, Liu2024}.

(ii)
Because the atoms move, the phase contribution $Kz$ in $\PL$ is Doppler shifted by $Kz\,v/c$.
Similarly, the FSL Doppler effect on the light phase $-\Delta\omega t$ gives rise to $-\Delta kz\,v/c$.
These contributions are usually~\cite{Dimopoulos2008, Tan2016, Tan2017, Liu2024,Tan2017relativistic} analyzed in the context of the correction term $3v_\pi/c$, where $v_\pi=v_0-gT+v_K/2$ is the mean velocity between both arms at the time of the mirror pulse and $v_K=\hbar K/\bar m$ is the recoil velocity.
In fact, the FSL Doppler effect arising from $\Delta \omega$ is usually suppressed in Bragg and Raman transitions, but it can be relevant in SPT and recoilless transitions as discussed below.

(iii)
The third contribution collects terms from chirping.
Even though chirping is not considered in all treatments~\cite{Dimopoulos2008}, the wave vector modification from $\delta\Phi$ gives rise to the term proportional to $\Delta k$ and is known to be amplified when not using opposite chirping~\cite{Cheng2015, Peters2001}.
In combination with the dominant FSL effect of chirping in $\Phi_\text L$, there are in total two chirp contributions.
They have been at the focus of different studies~\cite{Tan2016, Tan2017, Cheng2015}, where most articles define laser frequencies at the time of the mirror pulse instead at the time of the initial pulse.

(iv)
The mass defect in combination with the kinetic energy gives rise to a time-dilation contribution, which is very similar to the previous discussed FSL Doppler term.
To observe this contribution, the FSL Doppler effect of the atomic phase has to be included as well~\cite{Dimopoulos2008, Liu2024}.

For gravimetric applications, the gravitational acceleration is usually inferred from the chirping rate $\sigma$ when operating at the zero fringe where $\phi_{\text{AI}} =0$.
Since without perturbations we have $\phi_\text{AI}=-K(g-\sigma)T^2$, the gravitational acceleration is given by $g=\sigma$ for all interrogation times $T$.
Including relativistic effects, the position of the zero fringe is slightly shifted and depends on $T$.
Thus, the value of the chirp rate $\sigma$ at this point  does not coincide exactly with the gravitational acceleration $g$. 
We therefore solve $\phi_\text{AI}=0$ for $g$ and find an offset $\gamma$ from the idealized case through
\begin{equation}
\label{eq:chirp_gamma}
    g / \sigma = 1 + \gamma.
\end{equation}
By knowing the analytic form of the offset $\gamma$, the gravitational acceleration can be inferred from the chirp rate.
Explicit forms of the offset $\gamma$ are discussed for different diffraction mechanisms below.

\subsection{Single-photon transitions}
\label{subsec:SPT}
\begin{figure}[ht]
    \centering
    \includegraphics[page=4]{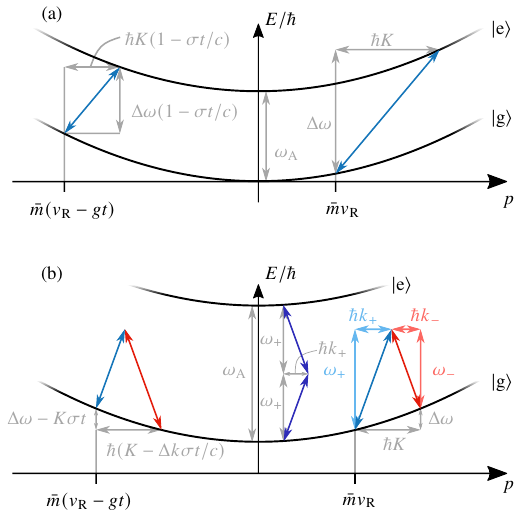}
    \caption{
        Energy-momentum diagrams for SPTs (a) and TPTs (b) are shown for the initial resonant velocity $v_\text R$ (right side) and a Doppler-shifted momentum including chirping (left side).
        The parabolas represent ground $\ket{\text{g}}$ and excited states $\ket{\text{e}}$ with an energy splitting $\omega_\text A$.
        The colored arrows show the driven transitions and the associated transferred energy and momenta, which are adjusted by chirping if the pulse is applied at a later time.
        Two-photon Bragg transitions (b) consist of sequential absorption and emission processes of counterpropagating light beams (blue and red), transferring a momentum of $\hbar K= \hbar (k_++k_-)$ and energy $\hbar \Delta \omega=\hbar (\omega_+-\omega_-)$.
        \mbox{E1-M1} recoilless transitions (purple, center) change the internal state by either absorbing or emitting one photon from each of two counterpropagating beams of identical frequency, transferring the energy $\hbar \omega_\text{A}$ without any momentum transfer.
        }
    \label{fig:parabola}
\end{figure}
SPTs at optical frequencies gain increasing attention for atom interferometry~\cite{Hu2017,Hu2019,Rudolph2020}, in particular for large-baseline~\cite{Abend2024,Abdalla2025,DiPumpo2024} gravitational-wave~\cite{Badurina2025} and dark-matter~\cite{Arvanitaki2018,Derr2023} detectors due to the suppression~\cite{Graham2013} of differential laser phase noise~\cite{LeGouet2007,Boehringer2025}.
In this case, only a single laser is used ($\Phi_-\equiv0$), so that $K=\Delta k$ and a single laser frequency determines the transferred energy $\Delta \omega = cK$, such that the transferred energy $\hbar \Delta \omega$ and momentum $\hbar K$ are connected by the dispersion relation of light.
Due to energy-momentum conservation, this laser frequency has to transfer the atomic energy $\hbar \omega_\text{A}$ and kinetic energy gained during the recoil for a resonant transition, see Fig.~\ref{fig:parabola}\,(a) for the energy-momentum diagram.
The right side shows a resonant transition at $t=0=z$.
However, the atom not only interacts with a single pulse but with multiple pulses at different times.
Since the atom is accelerated in between two pulses by gravity, changing the atom's momentum.
The left side of Fig.~\ref{fig:parabola}\,(a) shows the situation after some propagation time $t$.
To be still on resonance, the laser frequency has to be adjusted, which is achieved by chirping. 
Due to the dispersion relation in SPTs, this chirp implies also a change in the transferred momentum~\cite{Bott2023,Roura2025} that can be prominent.

Using Tables~\ref{tab:Disp} and \ref{tab:PhaseContr}, the total MZI phase for SPTs is given by
\begin{align}
    \phi_\text{S}=&
    -(k_\text{A}g-K\sigma)T^2-k_\text{A}gT^2 \, v_\pi/c.
\end{align}
The gravimetric phase $- k_\text{A} g T^2$ scales with the atomic wave vector $k_\text A$ as pointed out before~\cite{Dimopoulos2008, Graham2013, Rudolph2020,Liu2024} instead of scaling with the effective wave vector $K$.
In our framework, this effect is caused by FSL, while other treatments rely on different references frames~\cite{Badurina2025}.
However, because the unperturbed phase from chirping~\cite{Hu2019} still scales with $K$, the interferometer phase is subject to laser phase noise~\cite{Boehringer2025}.

Due to experimental imperfections, the initial velocity $v_0$ of the atoms is not known exactly.
Therefore, choosing a laser frequency defines a resonant velocity $v_\text R$, which may not necessarily coincide with $v_0$.
With the resonance condition $k_\text{A}= K [1 - v_\text{R} /c- v_\text K/(2c)]$ from Table~\ref{tab:Disp}, the SPT phase is then given in lowest order by
\begin{equation}
    \label{eq:PhaseSPTRes}
    \phi_\text{S} \cong -K(g-\sigma)T^2+KgT^2 (v_\text{R} - v_0 + g T)/c.
\end{equation}
In this case, the phase scales with $K$ and the resonance condition $v_\text R-v_0$ enters explicitly instead of the mean velocity $v_\pi$ at the mirror pulse.
Solving $\phi_\text S=0$ for $g$, we find the offset 
\begin{align}\label{eq:gammaS}
    \gamma_\text S= (v_\text r-v_0)/c + \sigma T/c
\end{align}
at the zero fringe.
It is purely determined by the deviation from resonance and chirping itself.
Even on resonance ($v_\text R=v_0$) this offset persists, which is a key difference to TPTs.

If the offset is accounted for in the estimation of $g$, it will introduce no systematic error to the accuracy of the measurement, but it may have an impact on the precision.
For that, we consider a random error caused by fluctuations $\Delta v_0$ of the initial velocity.
Assuming a phase uncertainty $\Delta\phi_\text S$ of the measurement, the the uncertainty $\Delta g$ of the measured gravitational acceleration is given by
\begin{align}\label{eq:phiS}
    \biggl(\frac{\Delta g}\sigma\biggr)^2=&
        \biggl(1+2\frac{v_\text r-v_0+2\sigma T}c\biggr)
            \biggl(\frac{\Delta\phi_\text S}{K\sigma T^2}\biggr)^2
        +\biggl(\frac{\Delta v_0}c\biggr)^2
\end{align}
using Gaussian error propagation linearized in $1/c$, but keeping the leading order of $\Delta v_0$ and neglecting any other uncertainties.

To get a rough order of magnitude, we  assume sub-Doppler cooling ($\Delta v_0/c \lesssim v_K/c \sim 10^{-11}$) so that the achievable precision on Earth is $\Delta g \sim 10^{-10}\,\text{m}/\text{s}^2$ (or $\si{\num{10}\,\nano\Gal}$).
Up to this precision, the measurement uncertainty will be dominated by phase noise.
However, a precision below this level requires a more careful analysis of $\Delta v_0$.
This uncertainty can be reduced by independent measurements of $v_0$, which requires additional resources and measurement time.
But if it is the limiting factor for precision, it can be mitigated by a suitable compensation technique in analogy to tip-tilt mirrors for rotations or wave-vector modifications for gravity gradients.
In our case, the occurrence of the initial velocity in the phase can be compensated by delaying the time of the mirror pulse as discussed in Appendix~\ref{apx:comp} so that the effect of velocity fluctuations is suppressed.
In most experimental configurations, however, these effects do not represent the primary constraint on precision.

\subsection{Bragg diffraction}
\label{subsec:Bragg}
Two-photon Bragg diffraction does not change the internal state of the atom ($k_\text A\equiv0$) and therefore offers the possibility of large momentum transfer through higher-order diffraction~\cite{Mueller2008,Hartmann2020}.
Figure~\ref{fig:parabola}\,(b) shows a Bragg transition on resonance on the right at $t=0=z$, where a photon from the blue laser is absorbed, transferring its momentum, followed by the stimulated emission into the counterpropagating red laser beam, so that the atom recoils again in the same direction.
The total transferred momentum $\hbar K= \hbar (k_+ + k_-)$ corresponds to the sum of both wave vectors and is in the order of an optical transition, while the transferred energy corresponds to the difference $\hbar \Delta \omega$ of photon energies, which for Bragg is typically in the order of kilohertz.
As shown on the right of Fig.~\ref{fig:parabola}\,(b), the resonance condition reads $\Delta \omega = \omega_K+\omega_\text{D}$ and implies $\Delta k = K (v_\text{R} + v_\text K/2)/c$.
Chirping ensures that the atoms are diffracted resonantly even if the atoms are accelerated by gravity in between the pulses, see  Fig.~\ref{fig:parabola}\,(b) on the left.
In contrast to SPTs, the transferred energy and momentum are chirped differently so that only a slight change of the momentum transfer is observed, since it depends on $\Delta k$.
The features of Bragg diffraction are summarized in Table~\ref{tab:Disp}.

Using the expressions from Table~\ref{tab:PhaseContr}, the phase becomes
\begin{align}
    \phi_\text B=&
        -K(g-\sigma)T^2\left(1+\frac{3v_\pi}{c}\right)
        -K\sigma T^2\frac{v_\pi+gT}{c}
        +\Delta k g T^2,
\end{align}
where $v_\pi=v_0-gT+v_K/2$ is again the mean velocity at the mirror pulse.
Here we omitted the contribution $-\hbar\Delta k\sigma t/c$, which is of order $c^{-2}$ due to the resonance condition.
In comparison to SPTs, the wave vector $k_\text A$ vanishes and thus the dominant gravitational phase scales with the effective wave vector $K$ of the light field.
Inserting the resonance condition Eq.~\eqref{eq:res_cond}, the total phase yields
\begin{align}
    \phi_\text B=&
        -K(g-\sigma)T^2\left(1+\frac{2v_\pi-gT}{c}\right)
        + K g T^2 \frac{v_\text{R} - v_0}c.
        \label{eq:phi_B}
\end{align}

On resonance, \ie{} $v_0=v_\text{R}$, FSL introduces an additional scaling factor $1+(2v_\pi-gT)/c$ without shifting the zero fringe.
As above, solving $\phi_\text{B}=0$ for $g$, the offset
\begin{align}
    \gamma_\text{B}=(v_\text r-v_0)/c
\end{align}
to gravitational measurements depends only on the resonance condition following Eq.~\eqref{eq:chirp_gamma}.
In contrast to SPTs, the wave vector chirp is suppressed and thus does not lead to an additional offset.

\subsection{Raman diffraction}
\label{subsec:Raman}
The other major two-photon process used in atom interferometry is Raman diffraction, which like Bragg diffraction allows to transfer momenta corresponding to optical transitions, while the internal state of the atom changes typically between hyperfine states.
Hence, the transferred energy corresponds to frequencies in the order of gigahertz, while they are in the order of kilohertz for Bragg transitions.
Thus, while being orders of magnitude larger than for Bragg, the fraction $\Delta k/K\sim\num{e-5}$ is still suppressed compared to SPTs.
The phase is given by
\begin{align}
    \phi_\text R=\phi_\text{B} + \Delta k \left[ (g-\sigma ) 2 v_\pi/c + g\sigma T/c\right] T^2
\end{align}
with the Bragg phase $\phi_\text B$ from Eq.~\eqref{eq:phi_B}.
In the limit of $\Delta k/K\approx0$, the Raman phase coincides with the Bragg phase.
The Raman offset of a gravity measurement is given by
\begin{equation}
    \gamma_\text R= \frac{v_\text{R}-v_0}{c}+\frac{\Delta k}K\frac{\sigma T}c
\end{equation}
at the zero fringe.
It reduces to the SPT offset $\gamma_\text S$ in the case $\Delta k/K=1$ and to the Bragg offset $\gamma_\text B$ for $\Delta k/K=0$.
For typical Raman transitions, the chirp contribution to this offset is suppressed by a factor of $\Delta k/K\sim10^{-5}$ in comparison to SPTs.

\section{Butterfly interferometry}
\label{sec:beyondMZI}
Since the phase expression from Sec.~\ref{sec:Ufrecht} is applicable to geometries beyond MZIs, we are in the position to study a  butterfly  or figure-of-eight configuration~\cite{Clauser1988,Kleinert2015}.
Such a setup is insensitive to linear accelerations and therefore not suitable for gravimetric applications. 
It consists of a beam splitter at $t=0$ generating the superposition, a first mirror at $t=T$ redirecting the arms to each other, a non-interacting crossing of both arms at $t=2T$, a second mirror pulse at $t=3T$ again redirecting the arms, and a final beam splitter at $t=4T$ generating the interference.
In our framework, this geometry is represented by the choice $\dot\Lambda_1(t)=\delta(t)-\delta(t-T)+\delta(t-3T)-\delta(t-4T)$ and $\dot\Lambda_2(t)=\delta(t-T)-\delta(t-3T)$.
The resulting phase contributions are listed in Table~\ref{tab:PhaseBFI}.
\begin{table}[ht]
    \centering
    \caption{
    Unperturbed phase $\phi_\text{un}$, phase contributions from Eq.~\eqref{eq:second_varphi}, and the resulting phase differences for a butterfly interferometer in a linear potential.
    }
    \begin{tabular}{lccc}
        \hline\hline
        Term & Integrand & Phase contribution\\
        \hline
        Unperturbed &  $\bullet$ &
            $\displaystyle0$\\
        \hline
        FSL clock & $\displaystyle\dot\Lambda(k_\text{A}-\Delta k)z$&
           $\displaystyle0$\\
        FSL Doppler & $\displaystyle\dot\Lambda(K-\Delta k)z v/c$& 
           $\displaystyle(\Delta k-K)\,g T^2 6gT/c$\\
        Chirp & $\displaystyle\dot\Lambda(K-\Delta k)z \sigma t/c$& 
           $(K-\Delta k)g T^2\,6\sigma T/c$\\
        Time dilation & $\displaystyle-\Lambda k_\text{A}v^2/(2c)$&
            $\displaystyle-k_\text{A}gT^2\,2gT/c$\\
        \hline\hline
    \end{tabular}
    \label{tab:PhaseBFI}
\end{table}

The unperturbed phase and linear perturbations cancel as expected.
Neglecting any gravity gradients as above, the remaining phase reads
\begin{equation}
    \phi_\text{BF}=-(K-\Delta k)gT^2\,6(g-\sigma)T/c-\Delta kgT^2\,2gT/c
\end{equation}
for all diffraction processes and making use of the resonance condition Eq.~\eqref{eq:res_cond}.
The first term cancels exactly for SPTs, while it is only suppressed for Bragg and Raman transitions if $\sigma \cong g$.
Furthermore, the last term is small for typical Bragg and Raman transitions.
Butterfly geometries are mainly used for measurements of higher moments of the gravitational potential or resonant enhancement of periodic signals, but not as gravimeters.
However, the phase shift $\phi_\text{BF}$ still has to be accounted for in such applications.

\section{FSL test with Doppler-free transitions}
\label{sec:E1M1}
Recoilless E1-M1 transitions represent a third type of TPT~\cite{Alden2014,Janson2024}, where the atom is excited by absorbing one photon from each of two counterpropagating light beams of identical frequency, see Fig.~\ref{fig:parabola}\,(b) in the center.
To connect to our results from Sec.~\ref{sec:LightGrav}, we do not flip the sign of $\Phi_-$ but of $k_-$, $\omega_-$, and $\phi_-$.
Because the beams have identical frequency, we therefore find $\omega_-=-\omega_+$ and $k_-=-k_+$.
Consequently, no net momentum is transferred since $K=k_++k_-=0$.
The transferred energy $\hbar \Delta \omega = \hbar (\omega_+-\omega_-)=\hbar 2 \omega_+$ implies the resonance condition  $\omega_\text{A}= \Delta\omega = 2 \omega_+$, with no transfer of kinetic energy.
As a result the resonance is independent of the velocity, \ie{} Doppler free, eliminating the need for chirping and thus $\sigma =0$.
Following Table~\ref{tab:PhaseContr}, the phase of an MZI driven with E1-M1 transition reads
\begin{equation}
    \phi_\text{E1-M1}= k_\text{A} g T^2 2 v_\pi /c,
\end{equation}
where the specific conditions for E1-M1 transitions from Table~\ref{tab:Disp} have been used.
Because the setup is not suitable for gravimetry, the interferometric phase is given by a single term which also was observed for Raman diffraction and that depends on the initial velocity.
However, in contrast to Raman diffraction, this term cannot be removed by the resonance condition, because it is independent of the velocity of the atom, so that the initial condition can only be suppressed in a differential setup.

To design a test for FSL effects, we therefore propose to use double-Bragg diffraction~\cite{Giese2013,Kueber2016,Ahlers2016,Hartmann2020} first to transfer the opposite velocity $v_\text B$ to both arms and perform an MZI sequence with recoilless E1-M1 transitions with the same lasers.
Both interferometers can be driven resonantly, because the transitions are Doppler free.
In this case, the differential phase $k_\text{A} g T^2 4v_\text{B}/c$ becomes independent of any initial conditions and contains FSL effects.
Using the clock-transition of strontium separating the arms by $v_\text B=\num{10}\,\si{\milli\meter/\second}$ and an interrogation time of $T=\num1\,\si s$ leads to a differential phase of $10^{-2}$, which is measurable with current setups.

\section{Conclusions}
\label{sec:conc}
By perturbatively solving the propagation of light in gravity including chirping and applying the resonance condition, we have demonstrated that the impact of FSL effects in atom interferometry depends on the used diffraction mechanism.
For a consistent treatment, we have taken into account:
(i) The relativistic mass defect, including redshift and time dilation;
(ii) frequency chirping to maintain resonance;
(iii) FSL effects in combination with the light phase $\Phi_\text L$ and the atomic phase $\PC$ as well as the wave vector chirp;
(iv) dispersion relations like $\omega_\text A=ck_\text{A}$ for correct power counting;
and (v) the FSL Doppler effect acting on frequency components of $\Phi_\text L$ and $\PC$.
Using a perturbative operator approach, we have identified four contributions, namely the FSL clock shift, the FSL Doppler shift, chirp contributions and time dilation effects that encompass also redshift contributions.
Based on these insights, we propose a test for pure FSL effects in a differential geometry using recoilless E1-M1 transitions.

For an MZI gravimeter, we were able to show that the resonance condition leads to a suppression of the initial velocity, so that the atom's velocity at the mirror pulse is less critical than the precise tuning of the lasers for resonant diffraction.
In fact, when operating Bragg diffraction on resonance, there is no offset ($\gamma_\text{B}=0$) in the measurement of the gravitational acceleration at the zero fringe.
In contrast to an optical SPT, we find  an offset of $\gamma_\text S = \sigma T/c \sim10^{-8}$ for an interrogation time of $T\sim\si{\num{.3}\,\s}$ on Earth, implying an accuracy of $\si{\num{e-7}\,\m\per\square\s}$ ($\si{\num{10}\,\micro\Gal}$).
Compared to SPTs, the Raman offset is suppressed by a factor of typical order $|\Delta k/K|\sim10^{-5}$ on resonance.
Off resonance, we observe an additional offset that depends on $(v_\text R-v_0)/c$, the difference between resonant and initial velocity, but not on the mean velocity at the mirror pulse.
In any case, velocity fluctuations affect the phase and therefore reduce the precision of the interferometer.
However, they can be mitigated by intentionally introducing a timing asymmetry similar to other compensation techniques. 

Beyond applications in gravimetry, we calculated the relativistic phase correction to a butterfly geometry for both TPTs and SPTs and observe contributions that can be considered to enhance the accuracy when measuring higher moments of the gravitational potential.

Our phase expression holds for various geometries and describes the dominant relativistic effects, for which we have assumed that a transition between internal states and a momentum transfer is described by the same state-response functions.
For example, large-momentum-transfer caused by higher-order Bragg diffraction or sequential pulses is well included in our treatment.
Considering large-momentum-transfer techniques relying on changing the direction of the momentum transfer~\cite{Graham2013,Rudolph2020} makes the introduction of different response functions for internal states and the external degrees of freedom (center-of-mass momentum) necessary.
This aspect is also true for double diffraction~\cite{Hartmann2020} techniques or recoil geometries~\cite{Schelfhout2024} like Ramsey-Bord\'e interferometers.
Nevertheless, the introduction of a second response function in our framework is straightforward, however, this aspect was beyond the scope of the present study and may be the subject of further work.

\section{Acknowledgments}
We thank M. Tartler, F. Di Pumpo, A. Friedrich and the QUANTUS team for helpful discussions.
The QUANTUS project is supported by the German Space Agency at the German Aerospace Center (Deutsche Raumfahrtagentur im Deutschen Zentrum f\"ur Luft- und Raumfahrt, DLR) with funds provided by the Federal Ministry for Economic Affairs and Climate Action (Bundesministerium f\"ur Wirtschaft und Klimaschutz, BMWK) due to an enactment of the German Bundestag under Grant No. 50WM2450E (QUANTUS-VI).

\appendix
\section{Compensation method}
\label{apx:comp}
This appendix focuses on compensating initial conditions induced by FSL in MZI geometries.
Effectively, FSL constitutes a position- and arm-dependent shift of the time of interaction.
By manipulating the pulse timing, this effect can be partially compensated.
The objective of such a procedure is to suppress a dependence on the initial atomic  velocity, enhancing the precision as discussed in the main body of the paper.
For that, we shift the time of the mirror pulse $T\mapsto T+\delta T$ by $\delta T$.
The time shift (TS) adds to the time shift caused by FSL, so that we have $\Delta T\mapsto\Delta T+\delta T(1+v/c)$.
It results in an additional perturbation potential given by
\begin{equation}
    V_\text{cmp}=
        \pm\delta(t-T)\bigl[
            (\omega_\text A-\Delta \omega)(1+v/c)
            +K(v-\sigma t)
        \bigr]\delta T
\end{equation}
for path $1$ ($+$) and path $2$ ($-$).
The first term arises from the frequency parts of the atomic phase $\PC$ and the laser phase $\PL$.
The second term corresponds to the dominant FSL effect of the momentum transfer $Kz$ and the dominant chirp term $-K\sigma t^2/2$ term of the light phase $\PL$.
The resulting phases are listed in Table~\ref{tab:PhaseContr2} and give rise to three additional contributions: a dominant phase (TS clock), a velocity-dependent term (TS Doppler), and a term that depends on the chirp (TS chirp).
These contributions can be directly compared to their counterparts induced by FSL given in Table~\ref{tab:PhaseContr}.
\begin{table}[ht]
    \centering
    \caption{
    Phase contributions induced by introducing a time shift $\delta T$ of the mirror pulse in an MZI geometry.
    }
    \begin{tabular}{lccc}
        \hline\hline
        Term & Integrand & Phase contribution\\
        \hline
        TS clock & $\displaystyle\dot\Lambda(\omega_\text A-\Delta \omega)\delta T$&
            $2(\Delta\omega-\omega_\text A)\delta T$\\
        TS Doppler & $\displaystyle\dot\Lambda(K+k_\text{A}-\Delta k)v \delta T  $&
            $2(\Delta k-k_\text{A}-K) v_\pi \delta T $\\
        TS chirp & $\displaystyle - \dot\Lambda K \sigma t\delta T$&
            $2K\sigma T\delta T$\\
        \hline\hline
    \end{tabular}
    \label{tab:PhaseContr2}
\end{table}
Using the resonance condition from Eq.~\eqref{eq:res_cond}, the TS clock term only acts as a perturbation, while the TS Doppler term is proportional to $K$.

To compensate the initial conditions in SPTs, we choose  $\Delta T=-(g+\Gamma)T^2/(2c)$, where $\Gamma\ll g$ encodes deviations from $g$ constrained by the precision of the gravimetric measurement.
Using the resonance condition, the interferometer phase then takes the form
\begin{equation}
    \frac{\phi_S}{KgT^2}=
        -1 + \frac \sigma g + \frac{\sigma T}{c}
        +\frac{\Gamma}{g}\frac{v_0-v_\text R-(g-\sigma)T}c.
\end{equation}
Solving for $g$ at $\phi_\text{S}=0$, the phase offset with $\Gamma =0$ becomes
\begin{align}
     g/\sigma=1+\sigma T/c
\end{align}
and solely depends on the chirp, so that the initial conditions are perfectly compensated in contrast to the previous expression from Eq.~\eqref{eq:gammaS}.
Gaussian error propagation for $g$ then leads to
\begin{align}
    \biggl(\frac{\Delta g}\sigma\biggr)^2=&
        \biggl[1-2\frac{ (\Gamma-\sigma) T}c\biggr]
            \biggl(\frac{\Delta\phi_\text S}{K\sigma T^2}\biggr)^2
        +\biggl(\frac{\Gamma}\sigma\frac{\Delta v_0}c\biggr)^2,
\end{align}
where only the uncertainties of the atom interferometer phase $\Delta\phi_\text S$ and fluctuations $\Delta v_0$ of the initial velocity are considered.
Furthermore, the expression is linearized in $1/c$ and the leading order for $\Delta v_0$ is kept.
The impact of $\Delta v_0$ is suppressed by the additional factor $\Gamma/\sigma\ll 1$, which arises from the relative precision in the a priori unknown gravitational acceleration $g$.
Similar results are achieved for the other diffraction mechanisms, for example choosing $\delta T=-(3g+3\Gamma-2\sigma)T^2/(2c)$ for Bragg diffraction.
\bibliography{literature.bib}

%apsrev4-2.bst 2019-01-14 (MD) hand-edited version of apsrev4-1.bst
%Control: key (0)
%Control: author (8) initials jnrlst
%Control: editor formatted (1) identically to author
%Control: production of article title (0) allowed
%Control: page (0) single
%Control: year (1) truncated
%Control: production of eprint (0) enabled
\begin{thebibliography}{67}%
\makeatletter
\providecommand \@ifxundefined [1]{%
 \@ifx{#1\undefined}
}%
\providecommand \@ifnum [1]{%
 \ifnum #1\expandafter \@firstoftwo
 \else \expandafter \@secondoftwo
 \fi
}%
\providecommand \@ifx [1]{%
 \ifx #1\expandafter \@firstoftwo
 \else \expandafter \@secondoftwo
 \fi
}%
\providecommand \natexlab [1]{#1}%
\providecommand \enquote  [1]{``#1''}%
\providecommand \bibnamefont  [1]{#1}%
\providecommand \bibfnamefont [1]{#1}%
\providecommand \citenamefont [1]{#1}%
\providecommand \href@noop [0]{\@secondoftwo}%
\providecommand \href [0]{\begingroup \@sanitize@url \@href}%
\providecommand \@href[1]{\@@startlink{#1}\@@href}%
\providecommand \@@href[1]{\endgroup#1\@@endlink}%
\providecommand \@sanitize@url [0]{\catcode `\\12\catcode `\$12\catcode
  `\&12\catcode `\#12\catcode `\^12\catcode `\_12\catcode `\%12\relax}%
\providecommand \@@startlink[1]{}%
\providecommand \@@endlink[0]{}%
\providecommand \url  [0]{\begingroup\@sanitize@url \@url }%
\providecommand \@url [1]{\endgroup\@href {#1}{\urlprefix }}%
\providecommand \urlprefix  [0]{URL }%
\providecommand \Eprint [0]{\href }%
\providecommand \doibase [0]{https://doi.org/}%
\providecommand \selectlanguage [0]{\@gobble}%
\providecommand \bibinfo  [0]{\@secondoftwo}%
\providecommand \bibfield  [0]{\@secondoftwo}%
\providecommand \translation [1]{[#1]}%
\providecommand \BibitemOpen [0]{}%
\providecommand \bibitemStop [0]{}%
\providecommand \bibitemNoStop [0]{.\EOS\space}%
\providecommand \EOS [0]{\spacefactor3000\relax}%
\providecommand \BibitemShut  [1]{\csname bibitem#1\endcsname}%
\let\auto@bib@innerbib\@empty
%</preamble>
\bibitem [{\citenamefont {Bongs}\ \emph {et~al.}(2019)\citenamefont {Bongs},
  \citenamefont {Holynski}, \citenamefont {Vovrosh}, \citenamefont {Bouyer},
  \citenamefont {Condon}, \citenamefont {Rasel}, \citenamefont {Schubert},
  \citenamefont {Schleich},\ and\ \citenamefont {Roura}}]{Bongs2019}%
  \BibitemOpen
  \bibfield  {author} {\bibinfo {author} {\bibfnamefont {K.}~\bibnamefont
  {Bongs}}, \bibinfo {author} {\bibfnamefont {M.}~\bibnamefont {Holynski}},
  \bibinfo {author} {\bibfnamefont {J.}~\bibnamefont {Vovrosh}}, \bibinfo
  {author} {\bibfnamefont {P.}~\bibnamefont {Bouyer}}, \bibinfo {author}
  {\bibfnamefont {G.}~\bibnamefont {Condon}}, \bibinfo {author} {\bibfnamefont
  {E.~M.}\ \bibnamefont {Rasel}}, \bibinfo {author} {\bibfnamefont
  {C.}~\bibnamefont {Schubert}}, \bibinfo {author} {\bibfnamefont {W.~P.}\
  \bibnamefont {Schleich}},\ and\ \bibinfo {author} {\bibfnamefont
  {A.}~\bibnamefont {Roura}},\ }\bibfield  {title} {\bibinfo {title} {Taking
  atom interferometric quantum sensors from the laboratory to real-world
  applications},\ }\href {https://doi.org/10.1038/s42254-019-0117-4} {\bibfield
   {journal} {\bibinfo  {journal} {Nat. Rev. Phys.}\ }\textbf {\bibinfo
  {volume} {1}},\ \bibinfo {pages} {731} (\bibinfo {year} {2019})}\BibitemShut
  {NoStop}%
\bibitem [{\citenamefont {Kovachy}\ \emph {et~al.}(2015)\citenamefont
  {Kovachy}, \citenamefont {Asenbaum}, \citenamefont {Overstreet},
  \citenamefont {Donnelly}, \citenamefont {Dickerson}, \citenamefont
  {Sugarbaker}, \citenamefont {Hogan},\ and\ \citenamefont
  {Kasevich}}]{Kovachy2015}%
  \BibitemOpen
  \bibfield  {author} {\bibinfo {author} {\bibfnamefont {T.}~\bibnamefont
  {Kovachy}}, \bibinfo {author} {\bibfnamefont {P.}~\bibnamefont {Asenbaum}},
  \bibinfo {author} {\bibfnamefont {C.}~\bibnamefont {Overstreet}}, \bibinfo
  {author} {\bibfnamefont {C.~A.}\ \bibnamefont {Donnelly}}, \bibinfo {author}
  {\bibfnamefont {S.~M.}\ \bibnamefont {Dickerson}}, \bibinfo {author}
  {\bibfnamefont {A.}~\bibnamefont {Sugarbaker}}, \bibinfo {author}
  {\bibfnamefont {J.~M.}\ \bibnamefont {Hogan}},\ and\ \bibinfo {author}
  {\bibfnamefont {M.~A.}\ \bibnamefont {Kasevich}},\ }\bibfield  {title}
  {\bibinfo {title} {Quantum superposition at the half-metre scale},\ }\href
  {https://doi.org/10.1038/nature16155} {\bibfield  {journal} {\bibinfo
  {journal} {Nature}\ }\textbf {\bibinfo {volume} {528}},\ \bibinfo {pages}
  {530} (\bibinfo {year} {2015})}\BibitemShut {NoStop}%
\bibitem [{\citenamefont {Abe}\ \emph {et~al.}(2021)\citenamefont {Abe},
  \citenamefont {Adamson}, \citenamefont {Borcean}, \citenamefont {Bortoletto},
  \citenamefont {Bridges}, \citenamefont {Carman}, \citenamefont
  {Chattopadhyay}, \citenamefont {Coleman}, \citenamefont {Curfman},
  \citenamefont {DeRose} \emph {et~al.}}]{Abe2021}%
  \BibitemOpen
  \bibfield  {author} {\bibinfo {author} {\bibfnamefont {M.}~\bibnamefont
  {Abe}}, \bibinfo {author} {\bibfnamefont {P.}~\bibnamefont {Adamson}},
  \bibinfo {author} {\bibfnamefont {M.}~\bibnamefont {Borcean}}, \bibinfo
  {author} {\bibfnamefont {D.}~\bibnamefont {Bortoletto}}, \bibinfo {author}
  {\bibfnamefont {K.}~\bibnamefont {Bridges}}, \bibinfo {author} {\bibfnamefont
  {S.~P.}\ \bibnamefont {Carman}}, \bibinfo {author} {\bibfnamefont
  {S.}~\bibnamefont {Chattopadhyay}}, \bibinfo {author} {\bibfnamefont
  {J.}~\bibnamefont {Coleman}}, \bibinfo {author} {\bibfnamefont {N.~M.}\
  \bibnamefont {Curfman}}, \bibinfo {author} {\bibfnamefont {K.}~\bibnamefont
  {DeRose}}, \emph {et~al.},\ }\bibfield  {title} {\bibinfo {title}
  {Matter-wave {{Atomic Gradiometer Interferometric Sensor}} ({{MAGIS-100}})},\
  }\href {https://doi.org/10.1088/2058-9565/abf719} {\bibfield  {journal}
  {\bibinfo  {journal} {Quantum Sci. Technol.}\ }\textbf {\bibinfo {volume}
  {6}},\ \bibinfo {pages} {044003} (\bibinfo {year} {2021})}\BibitemShut
  {NoStop}%
\bibitem [{\citenamefont {Debavelaere}\ \emph {et~al.}(2024)\citenamefont
  {Debavelaere}, \citenamefont {Solaro}, \citenamefont {Guellati-Kh\'elifa},\
  and\ \citenamefont {Clad\'e}}]{Debavelaere2024}%
  \BibitemOpen
  \bibfield  {author} {\bibinfo {author} {\bibfnamefont {C.}~\bibnamefont
  {Debavelaere}}, \bibinfo {author} {\bibfnamefont {C.}~\bibnamefont {Solaro}},
  \bibinfo {author} {\bibfnamefont {S.}~\bibnamefont {Guellati-Kh\'elifa}},\
  and\ \bibinfo {author} {\bibfnamefont {P.}~\bibnamefont {Clad\'e}},\
  }\bibfield  {title} {\bibinfo {title} {Atom interferometer using spatially
  localized beam splitters},\ }\href
  {https://doi.org/10.1103/PhysRevA.110.013310} {\bibfield  {journal} {\bibinfo
   {journal} {Phys. Rev. A}\ }\textbf {\bibinfo {volume} {110}},\ \bibinfo
  {pages} {013310} (\bibinfo {year} {2024})}\BibitemShut {NoStop}%
\bibitem [{\citenamefont {Dimopoulos}\ \emph {et~al.}(2008)\citenamefont
  {Dimopoulos}, \citenamefont {Graham}, \citenamefont {Hogan},\ and\
  \citenamefont {Kasevich}}]{Dimopoulos2008}%
  \BibitemOpen
  \bibfield  {author} {\bibinfo {author} {\bibfnamefont {S.}~\bibnamefont
  {Dimopoulos}}, \bibinfo {author} {\bibfnamefont {P.~W.}\ \bibnamefont
  {Graham}}, \bibinfo {author} {\bibfnamefont {J.~M.}\ \bibnamefont {Hogan}},\
  and\ \bibinfo {author} {\bibfnamefont {M.~A.}\ \bibnamefont {Kasevich}},\
  }\bibfield  {title} {\bibinfo {title} {General relativistic effects in atom
  interferometry},\ }\href {https://doi.org/10.1103/PhysRevD.78.042003}
  {\bibfield  {journal} {\bibinfo  {journal} {Phys. Rev. D}\ }\textbf {\bibinfo
  {volume} {78}},\ \bibinfo {pages} {042003} (\bibinfo {year}
  {2008})}\BibitemShut {NoStop}%
\bibitem [{\citenamefont {Liu}\ \emph {et~al.}(2024)\citenamefont {Liu},
  \citenamefont {Xu}, \citenamefont {Luo}, \citenamefont {Cao}, \citenamefont
  {Zhou}, \citenamefont {Duan},\ and\ \citenamefont {Hu}}]{Liu2024}%
  \BibitemOpen
  \bibfield  {author} {\bibinfo {author} {\bibfnamefont {J.}~\bibnamefont
  {Liu}}, \bibinfo {author} {\bibfnamefont {Y.}~\bibnamefont {Xu}}, \bibinfo
  {author} {\bibfnamefont {H.}~\bibnamefont {Luo}}, \bibinfo {author}
  {\bibfnamefont {L.}~\bibnamefont {Cao}}, \bibinfo {author} {\bibfnamefont
  {M.}~\bibnamefont {Zhou}}, \bibinfo {author} {\bibfnamefont {X.}~\bibnamefont
  {Duan}},\ and\ \bibinfo {author} {\bibfnamefont {Z.}~\bibnamefont {Hu}},\
  }\bibfield  {title} {\bibinfo {title} {Test of the gravitational redshift
  with single-photon-based atomic clock interferometers},\ }\href
  {https://doi.org/10.1007/s44214-024-00049-1} {\bibfield  {journal} {\bibinfo
  {journal} {Quantum Front.}\ }\textbf {\bibinfo {volume} {3}},\ \bibinfo
  {pages} {2} (\bibinfo {year} {2024})}\BibitemShut {NoStop}%
\bibitem [{\citenamefont {Hu}\ \emph {et~al.}(2017)\citenamefont {Hu},
  \citenamefont {Poli}, \citenamefont {Salvi},\ and\ \citenamefont
  {Tino}}]{Hu2017}%
  \BibitemOpen
  \bibfield  {author} {\bibinfo {author} {\bibfnamefont {L.}~\bibnamefont
  {Hu}}, \bibinfo {author} {\bibfnamefont {N.}~\bibnamefont {Poli}}, \bibinfo
  {author} {\bibfnamefont {L.}~\bibnamefont {Salvi}},\ and\ \bibinfo {author}
  {\bibfnamefont {G.~M.}\ \bibnamefont {Tino}},\ }\bibfield  {title} {\bibinfo
  {title} {Atom interferometry with the {Sr} optical clock transition},\ }\href
  {https://doi.org/10.1103/PhysRevLett.119.263601} {\bibfield  {journal}
  {\bibinfo  {journal} {Phys. Rev. Lett.}\ }\textbf {\bibinfo {volume} {119}},\
  \bibinfo {pages} {263601} (\bibinfo {year} {2017})}\BibitemShut {NoStop}%
\bibitem [{\citenamefont {Hu}\ \emph {et~al.}(2019)\citenamefont {Hu},
  \citenamefont {Wang}, \citenamefont {Salvi}, \citenamefont {Tinsley},
  \citenamefont {Tino},\ and\ \citenamefont {Poli}}]{Hu2019}%
  \BibitemOpen
  \bibfield  {author} {\bibinfo {author} {\bibfnamefont {L.}~\bibnamefont
  {Hu}}, \bibinfo {author} {\bibfnamefont {E.}~\bibnamefont {Wang}}, \bibinfo
  {author} {\bibfnamefont {L.}~\bibnamefont {Salvi}}, \bibinfo {author}
  {\bibfnamefont {J.~N.}\ \bibnamefont {Tinsley}}, \bibinfo {author}
  {\bibfnamefont {G.~M.}\ \bibnamefont {Tino}},\ and\ \bibinfo {author}
  {\bibfnamefont {N.}~\bibnamefont {Poli}},\ }\bibfield  {title} {\bibinfo
  {title} {Sr atom interferometry with the optical clock transition as a
  gravimeter and a gravity gradiometer},\ }\href
  {https://doi.org/10.1088/1361-6382/ab4d18} {\bibfield  {journal} {\bibinfo
  {journal} {Classical Quantum Gravity}\ }\textbf {\bibinfo {volume} {37}},\
  \bibinfo {pages} {014001} (\bibinfo {year} {2019})}\BibitemShut {NoStop}%
\bibitem [{\citenamefont {Rudolph}\ \emph {et~al.}(2020)\citenamefont
  {Rudolph}, \citenamefont {Wilkason}, \citenamefont {Nantel}, \citenamefont
  {Swan}, \citenamefont {Holland}, \citenamefont {Jiang}, \citenamefont
  {Garber}, \citenamefont {Carman},\ and\ \citenamefont {Hogan}}]{Rudolph2020}%
  \BibitemOpen
  \bibfield  {author} {\bibinfo {author} {\bibfnamefont {J.}~\bibnamefont
  {Rudolph}}, \bibinfo {author} {\bibfnamefont {T.}~\bibnamefont {Wilkason}},
  \bibinfo {author} {\bibfnamefont {M.}~\bibnamefont {Nantel}}, \bibinfo
  {author} {\bibfnamefont {H.}~\bibnamefont {Swan}}, \bibinfo {author}
  {\bibfnamefont {C.~M.}\ \bibnamefont {Holland}}, \bibinfo {author}
  {\bibfnamefont {Y.}~\bibnamefont {Jiang}}, \bibinfo {author} {\bibfnamefont
  {B.~E.}\ \bibnamefont {Garber}}, \bibinfo {author} {\bibfnamefont {S.~P.}\
  \bibnamefont {Carman}},\ and\ \bibinfo {author} {\bibfnamefont {J.~M.}\
  \bibnamefont {Hogan}},\ }\bibfield  {title} {\bibinfo {title} {Large
  {{Momentum Transfer Clock Atom Interferometry}} on the 689 nm
  {{Intercombination Line}} of {{Strontium}}},\ }\href
  {https://doi.org/10.1103/PhysRevLett.124.083604} {\bibfield  {journal}
  {\bibinfo  {journal} {Phys. Rev. Lett.}\ }\textbf {\bibinfo {volume} {124}},\
  \bibinfo {pages} {083604} (\bibinfo {year} {2020})}\BibitemShut {NoStop}%
\bibitem [{\citenamefont {Abend}\ \emph {et~al.}(2024)\citenamefont {Abend},
  \citenamefont {Allard}, \citenamefont {Alonso}, \citenamefont {Antoniadis},
  \citenamefont {Ara\'ujo}, \citenamefont {Arduini}, \citenamefont {Arnold},
  \citenamefont {Asano}, \citenamefont {Augst}, \citenamefont {Badurina} \emph
  {et~al.}}]{Abend2024}%
  \BibitemOpen
  \bibfield  {author} {\bibinfo {author} {\bibfnamefont {S.}~\bibnamefont
  {Abend}}, \bibinfo {author} {\bibfnamefont {B.}~\bibnamefont {Allard}},
  \bibinfo {author} {\bibfnamefont {I.}~\bibnamefont {Alonso}}, \bibinfo
  {author} {\bibfnamefont {J.}~\bibnamefont {Antoniadis}}, \bibinfo {author}
  {\bibfnamefont {H.}~\bibnamefont {Ara\'ujo}}, \bibinfo {author}
  {\bibfnamefont {G.}~\bibnamefont {Arduini}}, \bibinfo {author} {\bibfnamefont
  {A.~S.}\ \bibnamefont {Arnold}}, \bibinfo {author} {\bibfnamefont
  {T.}~\bibnamefont {Asano}}, \bibinfo {author} {\bibfnamefont
  {N.}~\bibnamefont {Augst}}, \bibinfo {author} {\bibfnamefont
  {L.}~\bibnamefont {Badurina}}, \emph {et~al.},\ }\bibfield  {title} {\bibinfo
  {title} {Terrestrial very-long-baseline atom interferometry: Workshop
  summary},\ }\href {https://doi.org/10.1116/5.0185291} {\bibfield  {journal}
  {\bibinfo  {journal} {AVS Quantum Sci.}\ }\textbf {\bibinfo {volume} {6}},\
  \bibinfo {pages} {024701} (\bibinfo {year} {2024})}\BibitemShut {NoStop}%
\bibitem [{\citenamefont {Abdalla}\ \emph {et~al.}(2025)\citenamefont
  {Abdalla}, \citenamefont {Abe}, \citenamefont {Abend}, \citenamefont {Abidi},
  \citenamefont {Aidelsburger}, \citenamefont {Alibabaei}, \citenamefont
  {Allard}, \citenamefont {Anoniadis}, \citenamefont {Arduini}, \citenamefont
  {Augst} \emph {et~al.}}]{Abdalla2025}%
  \BibitemOpen
  \bibfield  {author} {\bibinfo {author} {\bibfnamefont {A.}~\bibnamefont
  {Abdalla}}, \bibinfo {author} {\bibfnamefont {M.}~\bibnamefont {Abe}},
  \bibinfo {author} {\bibfnamefont {S.}~\bibnamefont {Abend}}, \bibinfo
  {author} {\bibfnamefont {M.}~\bibnamefont {Abidi}}, \bibinfo {author}
  {\bibfnamefont {M.}~\bibnamefont {Aidelsburger}}, \bibinfo {author}
  {\bibfnamefont {A.}~\bibnamefont {Alibabaei}}, \bibinfo {author}
  {\bibfnamefont {B.}~\bibnamefont {Allard}}, \bibinfo {author} {\bibfnamefont
  {J.}~\bibnamefont {Anoniadis}}, \bibinfo {author} {\bibfnamefont
  {G.}~\bibnamefont {Arduini}}, \bibinfo {author} {\bibfnamefont
  {N.}~\bibnamefont {Augst}}, \emph {et~al.},\ }\bibfield  {title} {\bibinfo
  {title} {Terrestrial very-long-baseline atom interferometry: summary of the
  second workshop},\ }\href {https://doi.org/10.1140/epjqt/s40507-025-00344-3}
  {\bibfield  {journal} {\bibinfo  {journal} {EPJ Quantum Technol.}\ }\textbf
  {\bibinfo {volume} {12}},\ \bibinfo {pages} {42} (\bibinfo {year}
  {2025})}\BibitemShut {NoStop}%
\bibitem [{\citenamefont {Di~Pumpo}\ \emph {et~al.}(2024)\citenamefont
  {Di~Pumpo}, \citenamefont {Friedrich},\ and\ \citenamefont
  {Giese}}]{DiPumpo2024}%
  \BibitemOpen
  \bibfield  {author} {\bibinfo {author} {\bibfnamefont {F.}~\bibnamefont
  {Di~Pumpo}}, \bibinfo {author} {\bibfnamefont {A.}~\bibnamefont
  {Friedrich}},\ and\ \bibinfo {author} {\bibfnamefont {E.}~\bibnamefont
  {Giese}},\ }\bibfield  {title} {\bibinfo {title} {Optimal baseline
  exploitation in vertical dark-matter detectors based on atom
  interferometry},\ }\href {https://doi.org/10.1116/5.0175683} {\bibfield
  {journal} {\bibinfo  {journal} {AVS Quantum Sci.}\ }\textbf {\bibinfo
  {volume} {6}},\ \bibinfo {pages} {014404} (\bibinfo {year}
  {2024})}\BibitemShut {NoStop}%
\bibitem [{\citenamefont {Graham}\ \emph {et~al.}(2013)\citenamefont {Graham},
  \citenamefont {Hogan}, \citenamefont {Kasevich},\ and\ \citenamefont
  {Rajendran}}]{Graham2013}%
  \BibitemOpen
  \bibfield  {author} {\bibinfo {author} {\bibfnamefont {P.~W.}\ \bibnamefont
  {Graham}}, \bibinfo {author} {\bibfnamefont {J.~M.}\ \bibnamefont {Hogan}},
  \bibinfo {author} {\bibfnamefont {M.~A.}\ \bibnamefont {Kasevich}},\ and\
  \bibinfo {author} {\bibfnamefont {S.}~\bibnamefont {Rajendran}},\ }\bibfield
  {title} {\bibinfo {title} {New {{Method}} for {{Gravitational Wave
  Detection}} with {{Atomic Sensors}}},\ }\href
  {https://doi.org/10.1103/PhysRevLett.110.171102} {\bibfield  {journal}
  {\bibinfo  {journal} {Phys. Rev. Lett.}\ }\textbf {\bibinfo {volume} {110}},\
  \bibinfo {pages} {171102} (\bibinfo {year} {2013})}\BibitemShut {NoStop}%
\bibitem [{\citenamefont {Badurina}\ \emph {et~al.}(2025)\citenamefont
  {Badurina}, \citenamefont {Du}, \citenamefont {Lee}, \citenamefont {Wang},\
  and\ \citenamefont {Zurek}}]{Badurina2025}%
  \BibitemOpen
  \bibfield  {author} {\bibinfo {author} {\bibfnamefont {L.}~\bibnamefont
  {Badurina}}, \bibinfo {author} {\bibfnamefont {Y.}~\bibnamefont {Du}},
  \bibinfo {author} {\bibfnamefont {V.~S.~H.}\ \bibnamefont {Lee}}, \bibinfo
  {author} {\bibfnamefont {Y.}~\bibnamefont {Wang}},\ and\ \bibinfo {author}
  {\bibfnamefont {K.~M.}\ \bibnamefont {Zurek}},\ }\bibfield  {title} {\bibinfo
  {title} {Signatures of linearized gravity in atom interferometers: A
  simplified computational framework},\ }\href
  {https://doi.org/10.1103/PhysRevD.111.042002} {\bibfield  {journal} {\bibinfo
   {journal} {Phys. Rev. D}\ }\textbf {\bibinfo {volume} {111}},\ \bibinfo
  {pages} {042002} (\bibinfo {year} {2025})}\BibitemShut {NoStop}%
\bibitem [{\citenamefont {Arvanitaki}\ \emph {et~al.}(2018)\citenamefont
  {Arvanitaki}, \citenamefont {Graham}, \citenamefont {Hogan}, \citenamefont
  {Rajendran},\ and\ \citenamefont {van Tilburg}}]{Arvanitaki2018}%
  \BibitemOpen
  \bibfield  {author} {\bibinfo {author} {\bibfnamefont {A.}~\bibnamefont
  {Arvanitaki}}, \bibinfo {author} {\bibfnamefont {P.~W.}\ \bibnamefont
  {Graham}}, \bibinfo {author} {\bibfnamefont {J.~M.}\ \bibnamefont {Hogan}},
  \bibinfo {author} {\bibfnamefont {S.}~\bibnamefont {Rajendran}},\ and\
  \bibinfo {author} {\bibfnamefont {K.}~\bibnamefont {van Tilburg}},\
  }\bibfield  {title} {\bibinfo {title} {Search for light scalar dark matter
  with atomic gravitational wave detectors},\ }\href
  {https://doi.org/10.1103/PhysRevD.97.075020} {\bibfield  {journal} {\bibinfo
  {journal} {Phys. Rev. D}\ }\textbf {\bibinfo {volume} {97}},\ \bibinfo
  {pages} {075020} (\bibinfo {year} {2018})}\BibitemShut {NoStop}%
\bibitem [{\citenamefont {Derr}\ and\ \citenamefont {Giese}(2023)}]{Derr2023}%
  \BibitemOpen
  \bibfield  {author} {\bibinfo {author} {\bibfnamefont {D.}~\bibnamefont
  {Derr}}\ and\ \bibinfo {author} {\bibfnamefont {E.}~\bibnamefont {Giese}},\
  }\bibfield  {title} {\bibinfo {title} {Clock transitions versus {Bragg}
  diffraction in atom-interferometric dark-matter detection},\ }\href
  {https://doi.org/10.1116/5.0176666} {\bibfield  {journal} {\bibinfo
  {journal} {AVS Quantum Sci.}\ }\textbf {\bibinfo {volume} {5}},\ \bibinfo
  {pages} {044404} (\bibinfo {year} {2023})}\BibitemShut {NoStop}%
\bibitem [{\citenamefont {Marzlin}\ and\ \citenamefont
  {Audretsch}(1996)}]{Marzlin1996}%
  \BibitemOpen
  \bibfield  {author} {\bibinfo {author} {\bibfnamefont {K.-P.}\ \bibnamefont
  {Marzlin}}\ and\ \bibinfo {author} {\bibfnamefont {J.}~\bibnamefont
  {Audretsch}},\ }\bibfield  {title} {\bibinfo {title} {``{{Freely}}'' falling
  two-level atom in a running laser wave},\ }\href
  {https://doi.org/10.1103/PhysRevA.53.1004} {\bibfield  {journal} {\bibinfo
  {journal} {Phys. Rev. A}\ }\textbf {\bibinfo {volume} {53}},\ \bibinfo
  {pages} {1004} (\bibinfo {year} {1996})}\BibitemShut {NoStop}%
\bibitem [{\citenamefont {Bott}\ \emph {et~al.}(2023)\citenamefont {Bott},
  \citenamefont {Di~Pumpo},\ and\ \citenamefont {Giese}}]{Bott2023}%
  \BibitemOpen
  \bibfield  {author} {\bibinfo {author} {\bibfnamefont {A.}~\bibnamefont
  {Bott}}, \bibinfo {author} {\bibfnamefont {F.}~\bibnamefont {Di~Pumpo}},\
  and\ \bibinfo {author} {\bibfnamefont {E.}~\bibnamefont {Giese}},\ }\bibfield
   {title} {\bibinfo {title} {Atomic diffraction from single-photon transitions
  in gravity and {{Standard-Model}} extensions},\ }\href
  {https://doi.org/10.1116/5.0174258} {\bibfield  {journal} {\bibinfo
  {journal} {AVS Quantum Sci.}\ }\textbf {\bibinfo {volume} {5}},\ \bibinfo
  {pages} {044402} (\bibinfo {year} {2023})}\BibitemShut {NoStop}%
\bibitem [{\citenamefont {B\"ohringer}\ and\ \citenamefont
  {Friedrich}(2024)}]{Boehringer2024}%
  \BibitemOpen
  \bibfield  {author} {\bibinfo {author} {\bibfnamefont {S.}~\bibnamefont
  {B\"ohringer}}\ and\ \bibinfo {author} {\bibfnamefont {A.}~\bibnamefont
  {Friedrich}},\ }\bibfield  {title} {\bibinfo {title} {Decoupling of external
  and internal dynamics in driven two-level systems},\ }\href
  {https://doi.org/10.1103/PhysRevResearch.6.043153} {\bibfield  {journal}
  {\bibinfo  {journal} {Phys. Rev. Res.}\ }\textbf {\bibinfo {volume} {6}},\
  \bibinfo {pages} {043153} (\bibinfo {year} {2024})}\BibitemShut {NoStop}%
\bibitem [{\citenamefont {Tan}\ \emph {et~al.}(2017{\natexlab{a}})\citenamefont
  {Tan}, \citenamefont {Shao},\ and\ \citenamefont {Hu}}]{Tan2017relativistic}%
  \BibitemOpen
  \bibfield  {author} {\bibinfo {author} {\bibfnamefont {Y.-J.}\ \bibnamefont
  {Tan}}, \bibinfo {author} {\bibfnamefont {C.-G.}\ \bibnamefont {Shao}},\ and\
  \bibinfo {author} {\bibfnamefont {Z.-K.}\ \bibnamefont {Hu}},\ }\bibfield
  {title} {\bibinfo {title} {Relativistic effects in atom gravimeters},\ }\href
  {https://doi.org/10.1103/PhysRevD.95.024002} {\bibfield  {journal} {\bibinfo
  {journal} {Phys. Rev. D}\ }\textbf {\bibinfo {volume} {95}},\ \bibinfo
  {pages} {024002} (\bibinfo {year} {2017}{\natexlab{a}})}\BibitemShut
  {NoStop}%
\bibitem [{\citenamefont {Di~Pumpo}\ \emph {et~al.}(2021)\citenamefont
  {Di~Pumpo}, \citenamefont {Ufrecht}, \citenamefont {Friedrich}, \citenamefont
  {Giese}, \citenamefont {Schleich},\ and\ \citenamefont
  {Unruh}}]{DiPumpo2021}%
  \BibitemOpen
  \bibfield  {author} {\bibinfo {author} {\bibfnamefont {F.}~\bibnamefont
  {Di~Pumpo}}, \bibinfo {author} {\bibfnamefont {C.}~\bibnamefont {Ufrecht}},
  \bibinfo {author} {\bibfnamefont {A.}~\bibnamefont {Friedrich}}, \bibinfo
  {author} {\bibfnamefont {E.}~\bibnamefont {Giese}}, \bibinfo {author}
  {\bibfnamefont {W.~P.}\ \bibnamefont {Schleich}},\ and\ \bibinfo {author}
  {\bibfnamefont {W.~G.}\ \bibnamefont {Unruh}},\ }\bibfield  {title} {\bibinfo
  {title} {Gravitational redshift tests with atomic clocks and atom
  interferometers},\ }\href {https://doi.org/10.1103/PRXQuantum.2.040333}
  {\bibfield  {journal} {\bibinfo  {journal} {PRX Quantum}\ }\textbf {\bibinfo
  {volume} {2}},\ \bibinfo {pages} {040333} (\bibinfo {year}
  {2021})}\BibitemShut {NoStop}%
\bibitem [{\citenamefont {Ufrecht}\ \emph {et~al.}(2020)\citenamefont
  {Ufrecht}, \citenamefont {Di~Pumpo}, \citenamefont {Friedrich}, \citenamefont
  {Roura}, \citenamefont {Schubert}, \citenamefont {Schlippert}, \citenamefont
  {Rasel}, \citenamefont {Schleich},\ and\ \citenamefont
  {Giese}}]{Ufrecht2020ugr}%
  \BibitemOpen
  \bibfield  {author} {\bibinfo {author} {\bibfnamefont {C.}~\bibnamefont
  {Ufrecht}}, \bibinfo {author} {\bibfnamefont {F.}~\bibnamefont {Di~Pumpo}},
  \bibinfo {author} {\bibfnamefont {A.}~\bibnamefont {Friedrich}}, \bibinfo
  {author} {\bibfnamefont {A.}~\bibnamefont {Roura}}, \bibinfo {author}
  {\bibfnamefont {C.}~\bibnamefont {Schubert}}, \bibinfo {author}
  {\bibfnamefont {D.}~\bibnamefont {Schlippert}}, \bibinfo {author}
  {\bibfnamefont {E.~M.}\ \bibnamefont {Rasel}}, \bibinfo {author}
  {\bibfnamefont {W.~P.}\ \bibnamefont {Schleich}},\ and\ \bibinfo {author}
  {\bibfnamefont {E.}~\bibnamefont {Giese}},\ }\bibfield  {title} {\bibinfo
  {title} {Atom-interferometric test of the universality of gravitational
  redshift and free fall},\ }\href
  {https://doi.org/10.1103/PhysRevResearch.2.043240} {\bibfield  {journal}
  {\bibinfo  {journal} {Phys. Rev. Res.}\ }\textbf {\bibinfo {volume} {2}},\
  \bibinfo {pages} {043240} (\bibinfo {year} {2020})}\BibitemShut {NoStop}%
\bibitem [{\citenamefont {Roura}(2020)}]{Roura2020}%
  \BibitemOpen
  \bibfield  {author} {\bibinfo {author} {\bibfnamefont {A.}~\bibnamefont
  {Roura}},\ }\bibfield  {title} {\bibinfo {title} {Gravitational redshift in
  quantum-clock interferometry},\ }\href
  {https://doi.org/10.1103/PhysRevX.10.021014} {\bibfield  {journal} {\bibinfo
  {journal} {Phys. Rev. X}\ }\textbf {\bibinfo {volume} {10}},\ \bibinfo
  {pages} {021014} (\bibinfo {year} {2020})}\BibitemShut {NoStop}%
\bibitem [{\citenamefont {Di~Pumpo}\ \emph {et~al.}(2023)\citenamefont
  {Di~Pumpo}, \citenamefont {Friedrich}, \citenamefont {Ufrecht},\ and\
  \citenamefont {Giese}}]{DiPumpo2023}%
  \BibitemOpen
  \bibfield  {author} {\bibinfo {author} {\bibfnamefont {F.}~\bibnamefont
  {Di~Pumpo}}, \bibinfo {author} {\bibfnamefont {A.}~\bibnamefont {Friedrich}},
  \bibinfo {author} {\bibfnamefont {C.}~\bibnamefont {Ufrecht}},\ and\ \bibinfo
  {author} {\bibfnamefont {E.}~\bibnamefont {Giese}},\ }\bibfield  {title}
  {\bibinfo {title} {Universality-of-clock-rates test using atom interferometry
  with ${T}^{3}$ scaling},\ }\href
  {https://doi.org/10.1103/PhysRevD.107.064007} {\bibfield  {journal} {\bibinfo
   {journal} {Phys. Rev. D}\ }\textbf {\bibinfo {volume} {107}},\ \bibinfo
  {pages} {064007} (\bibinfo {year} {2023})}\BibitemShut {NoStop}%
\bibitem [{\citenamefont {Loriani}\ \emph {et~al.}(2019)\citenamefont
  {Loriani}, \citenamefont {Friedrich}, \citenamefont {Ufrecht}, \citenamefont
  {Di~Pumpo}, \citenamefont {Kleinert}, \citenamefont {Abend}, \citenamefont
  {Gaaloul}, \citenamefont {Meiners}, \citenamefont {Schubert}, \citenamefont
  {Tell} \emph {et~al.}}]{Loriani2019}%
  \BibitemOpen
  \bibfield  {author} {\bibinfo {author} {\bibfnamefont {S.}~\bibnamefont
  {Loriani}}, \bibinfo {author} {\bibfnamefont {A.}~\bibnamefont {Friedrich}},
  \bibinfo {author} {\bibfnamefont {C.}~\bibnamefont {Ufrecht}}, \bibinfo
  {author} {\bibfnamefont {F.}~\bibnamefont {Di~Pumpo}}, \bibinfo {author}
  {\bibfnamefont {S.}~\bibnamefont {Kleinert}}, \bibinfo {author}
  {\bibfnamefont {S.}~\bibnamefont {Abend}}, \bibinfo {author} {\bibfnamefont
  {N.}~\bibnamefont {Gaaloul}}, \bibinfo {author} {\bibfnamefont
  {C.}~\bibnamefont {Meiners}}, \bibinfo {author} {\bibfnamefont
  {C.}~\bibnamefont {Schubert}}, \bibinfo {author} {\bibfnamefont
  {D.}~\bibnamefont {Tell}}, \emph {et~al.},\ }\bibfield  {title} {\bibinfo
  {title} {Interference of clocks: {A} quantum twin paradox},\ }\href
  {https://advances.sciencemag.org/content/5/10/eaax8966} {\bibfield  {journal}
  {\bibinfo  {journal} {Sci. Adv.}\ }\textbf {\bibinfo {volume} {5}},\ \bibinfo
  {pages} {eaax8966} (\bibinfo {year} {2019})}\BibitemShut {NoStop}%
\bibitem [{\citenamefont {Roura}\ \emph {et~al.}(2021)\citenamefont {Roura},
  \citenamefont {Schubert}, \citenamefont {Schlippert},\ and\ \citenamefont
  {Rasel}}]{Roura2021}%
  \BibitemOpen
  \bibfield  {author} {\bibinfo {author} {\bibfnamefont {A.}~\bibnamefont
  {Roura}}, \bibinfo {author} {\bibfnamefont {C.}~\bibnamefont {Schubert}},
  \bibinfo {author} {\bibfnamefont {D.}~\bibnamefont {Schlippert}},\ and\
  \bibinfo {author} {\bibfnamefont {E.~M.}\ \bibnamefont {Rasel}},\ }\bibfield
  {title} {\bibinfo {title} {Measuring gravitational time dilation with
  delocalized quantum superpositions},\ }\href
  {https://doi.org/10.1103/PhysRevD.104.084001} {\bibfield  {journal} {\bibinfo
   {journal} {Phys. Rev. D}\ }\textbf {\bibinfo {volume} {104}},\ \bibinfo
  {pages} {084001} (\bibinfo {year} {2021})}\BibitemShut {NoStop}%
\bibitem [{\citenamefont {Pikovski}\ \emph {et~al.}(2017)\citenamefont
  {Pikovski}, \citenamefont {Zych}, \citenamefont {Costa},\ and\ \citenamefont
  {Brukner}}]{Pikovski2017}%
  \BibitemOpen
  \bibfield  {author} {\bibinfo {author} {\bibfnamefont {I.}~\bibnamefont
  {Pikovski}}, \bibinfo {author} {\bibfnamefont {M.}~\bibnamefont {Zych}},
  \bibinfo {author} {\bibfnamefont {F.}~\bibnamefont {Costa}},\ and\ \bibinfo
  {author} {\bibfnamefont {{\v{C}}.}~\bibnamefont {Brukner}},\ }\bibfield
  {title} {\bibinfo {title} {Time dilation in quantum systems and
  decoherence},\ }\href {https://doi.org/10.1088/1367-2630/aa5d92} {\bibfield
  {journal} {\bibinfo  {journal} {New J. Phys.}\ }\textbf {\bibinfo {volume}
  {19}},\ \bibinfo {pages} {025011} (\bibinfo {year} {2017})}\BibitemShut
  {NoStop}%
\bibitem [{\citenamefont {Zych}\ \emph {et~al.}(2011)\citenamefont {Zych},
  \citenamefont {Costa}, \citenamefont {Pikovski},\ and\ \citenamefont
  {Brukner}}]{Zych2011}%
  \BibitemOpen
  \bibfield  {author} {\bibinfo {author} {\bibfnamefont {M.}~\bibnamefont
  {Zych}}, \bibinfo {author} {\bibfnamefont {F.}~\bibnamefont {Costa}},
  \bibinfo {author} {\bibfnamefont {I.}~\bibnamefont {Pikovski}},\ and\
  \bibinfo {author} {\bibfnamefont {{\v{C}}.}~\bibnamefont {Brukner}},\
  }\bibfield  {title} {\bibinfo {title} {Quantum interferometric visibility as
  a witness of general relativistic proper time},\ }\href
  {https://www.nature.com/articles/ncomms1498} {\bibfield  {journal} {\bibinfo
  {journal} {Nat. Commun.}\ }\textbf {\bibinfo {volume} {2}},\ \bibinfo {pages}
  {505} (\bibinfo {year} {2011})}\BibitemShut {NoStop}%
\bibitem [{\citenamefont {Hartmann}\ \emph {et~al.}(2020)\citenamefont
  {Hartmann}, \citenamefont {Jenewein}, \citenamefont {Giese}, \citenamefont
  {Abend}, \citenamefont {Roura}, \citenamefont {Rasel},\ and\ \citenamefont
  {Schleich}}]{Hartmann2020}%
  \BibitemOpen
  \bibfield  {author} {\bibinfo {author} {\bibfnamefont {S.}~\bibnamefont
  {Hartmann}}, \bibinfo {author} {\bibfnamefont {J.}~\bibnamefont {Jenewein}},
  \bibinfo {author} {\bibfnamefont {E.}~\bibnamefont {Giese}}, \bibinfo
  {author} {\bibfnamefont {S.}~\bibnamefont {Abend}}, \bibinfo {author}
  {\bibfnamefont {A.}~\bibnamefont {Roura}}, \bibinfo {author} {\bibfnamefont
  {E.~M.}\ \bibnamefont {Rasel}},\ and\ \bibinfo {author} {\bibfnamefont
  {W.~P.}\ \bibnamefont {Schleich}},\ }\bibfield  {title} {\bibinfo {title}
  {Regimes of atomic diffraction: {Raman} versus {Bragg} diffraction in
  retroreflective geometries},\ }\href
  {https://doi.org/10.1103/PhysRevA.101.053610} {\bibfield  {journal} {\bibinfo
   {journal} {Phys. Rev. A}\ }\textbf {\bibinfo {volume} {101}},\ \bibinfo
  {pages} {053610} (\bibinfo {year} {2020})}\BibitemShut {NoStop}%
\bibitem [{\citenamefont {Peters}\ \emph {et~al.}(2001)\citenamefont {Peters},
  \citenamefont {Chung},\ and\ \citenamefont {Chu}}]{Peters2001}%
  \BibitemOpen
  \bibfield  {author} {\bibinfo {author} {\bibfnamefont {A.}~\bibnamefont
  {Peters}}, \bibinfo {author} {\bibfnamefont {K.~Y.}\ \bibnamefont {Chung}},\
  and\ \bibinfo {author} {\bibfnamefont {S.}~\bibnamefont {Chu}},\ }\bibfield
  {title} {\bibinfo {title} {High-precision gravity measurements using atom
  interferometry},\ }\href {https://doi.org/10.1088/0026-1394/38/1/4}
  {\bibfield  {journal} {\bibinfo  {journal} {Metrologia}\ }\textbf {\bibinfo
  {volume} {38}},\ \bibinfo {pages} {25} (\bibinfo {year} {2001})}\BibitemShut
  {NoStop}%
\bibitem [{\citenamefont {Cheng}\ \emph {et~al.}(2015)\citenamefont {Cheng},
  \citenamefont {Gillot}, \citenamefont {Merlet},\ and\ \citenamefont {Pereira
  Dos~Santos}}]{Cheng2015}%
  \BibitemOpen
  \bibfield  {author} {\bibinfo {author} {\bibfnamefont {B.}~\bibnamefont
  {Cheng}}, \bibinfo {author} {\bibfnamefont {P.}~\bibnamefont {Gillot}},
  \bibinfo {author} {\bibfnamefont {S.}~\bibnamefont {Merlet}},\ and\ \bibinfo
  {author} {\bibfnamefont {F.}~\bibnamefont {Pereira Dos~Santos}},\ }\bibfield
  {title} {\bibinfo {title} {Influence of chirping the {{Raman}} lasers in an
  atom gravimeter: {{Phase}} shifts due to the {{Raman}} light shift and to the
  finite speed of light},\ }\href {https://doi.org/10.1103/PhysRevA.92.063617}
  {\bibfield  {journal} {\bibinfo  {journal} {Phys. Rev. A}\ }\textbf {\bibinfo
  {volume} {92}},\ \bibinfo {pages} {063617} (\bibinfo {year}
  {2015})}\BibitemShut {NoStop}%
\bibitem [{\citenamefont {Tan}\ \emph {et~al.}(2016)\citenamefont {Tan},
  \citenamefont {Shao},\ and\ \citenamefont {Hu}}]{Tan2016}%
  \BibitemOpen
  \bibfield  {author} {\bibinfo {author} {\bibfnamefont {Y.~J.}\ \bibnamefont
  {Tan}}, \bibinfo {author} {\bibfnamefont {C.~G.}\ \bibnamefont {Shao}},\ and\
  \bibinfo {author} {\bibfnamefont {Z.~K.}\ \bibnamefont {Hu}},\ }\bibfield
  {title} {\bibinfo {title} {{F}inite-speed-of-light perturbation in atom
  gravimeters},\ }\href {https://doi.org/10.1103/PhysRevA.94.013612} {\bibfield
   {journal} {\bibinfo  {journal} {Phys. Rev. A}\ }\textbf {\bibinfo {volume}
  {94}},\ \bibinfo {pages} {013612} (\bibinfo {year} {2016})}\BibitemShut
  {NoStop}%
\bibitem [{\citenamefont {Tan}\ \emph {et~al.}(2017{\natexlab{b}})\citenamefont
  {Tan}, \citenamefont {Shao},\ and\ \citenamefont {Hu}}]{Tan2017}%
  \BibitemOpen
  \bibfield  {author} {\bibinfo {author} {\bibfnamefont {Y.-J.}\ \bibnamefont
  {Tan}}, \bibinfo {author} {\bibfnamefont {C.-G.}\ \bibnamefont {Shao}},\ and\
  \bibinfo {author} {\bibfnamefont {Z.-K.}\ \bibnamefont {Hu}},\ }\bibfield
  {title} {\bibinfo {title} {Time delay and the effect of the finite speed of
  light in atom gravimeters},\ }\href
  {https://doi.org/10.1103/PhysRevA.96.023604} {\bibfield  {journal} {\bibinfo
  {journal} {Phys. Rev. A}\ }\textbf {\bibinfo {volume} {96}},\ \bibinfo
  {pages} {023604} (\bibinfo {year} {2017}{\natexlab{b}})}\BibitemShut
  {NoStop}%
\bibitem [{\citenamefont {Xu}\ \emph {et~al.}(2022)\citenamefont {Xu},
  \citenamefont {Deng}, \citenamefont {Duan}, \citenamefont {Cao},
  \citenamefont {Zhou}, \citenamefont {Shao},\ and\ \citenamefont
  {Hu}}]{Xu2022}%
  \BibitemOpen
  \bibfield  {author} {\bibinfo {author} {\bibfnamefont {Y.-Y.}\ \bibnamefont
  {Xu}}, \bibinfo {author} {\bibfnamefont {X.-B.}\ \bibnamefont {Deng}},
  \bibinfo {author} {\bibfnamefont {X.-C.}\ \bibnamefont {Duan}}, \bibinfo
  {author} {\bibfnamefont {L.-S.}\ \bibnamefont {Cao}}, \bibinfo {author}
  {\bibfnamefont {M.-K.}\ \bibnamefont {Zhou}}, \bibinfo {author}
  {\bibfnamefont {C.-G.}\ \bibnamefont {Shao}},\ and\ \bibinfo {author}
  {\bibfnamefont {Z.-K.}\ \bibnamefont {Hu}},\ }\bibfield  {title} {\bibinfo
  {title} {de-{Broglie} wavelength enhanced weak equivalence principle test for
  atoms in different hyperfine states},\ }\href
  {https://doi.org/10.48550/arXiv.2210.08533} {\bibfield  {journal} {\bibinfo
  {journal} {arXiv:2210.08533}\ } (\bibinfo {year} {2022})}\BibitemShut
  {NoStop}%
\bibitem [{\citenamefont {Asano}\ \emph {et~al.}(2024)\citenamefont {Asano},
  \citenamefont {Giese},\ and\ \citenamefont {Di~Pumpo}}]{Asano2024}%
  \BibitemOpen
  \bibfield  {author} {\bibinfo {author} {\bibfnamefont {T.}~\bibnamefont
  {Asano}}, \bibinfo {author} {\bibfnamefont {E.}~\bibnamefont {Giese}},\ and\
  \bibinfo {author} {\bibfnamefont {F.}~\bibnamefont {Di~Pumpo}},\ }\bibfield
  {title} {\bibinfo {title} {Quantum field theory for multipolar composite
  bosons with mass defect and relativistic corrections},\ }\href
  {https://doi.org/10.1103/PRXQuantum.5.020322} {\bibfield  {journal} {\bibinfo
   {journal} {PRX Quantum}\ }\textbf {\bibinfo {volume} {5}},\ \bibinfo {pages}
  {020322} (\bibinfo {year} {2024})}\BibitemShut {NoStop}%
\bibitem [{\citenamefont {Sonnleitner}\ and\ \citenamefont
  {Barnett}(2018)}]{Sonnleitner2018}%
  \BibitemOpen
  \bibfield  {author} {\bibinfo {author} {\bibfnamefont {M.}~\bibnamefont
  {Sonnleitner}}\ and\ \bibinfo {author} {\bibfnamefont {S.~M.}\ \bibnamefont
  {Barnett}},\ }\bibfield  {title} {\bibinfo {title} {Mass-energy and anomalous
  friction in quantum optics},\ }\href
  {https://doi.org/10.1103/PhysRevA.98.042106} {\bibfield  {journal} {\bibinfo
  {journal} {Phys. Rev. A}\ }\textbf {\bibinfo {volume} {98}},\ \bibinfo
  {pages} {042106} (\bibinfo {year} {2018})}\BibitemShut {NoStop}%
\bibitem [{\citenamefont {Schwartz}\ and\ \citenamefont
  {Giulini}(2019)}]{Schwartz2019}%
  \BibitemOpen
  \bibfield  {author} {\bibinfo {author} {\bibfnamefont {P.~K.}\ \bibnamefont
  {Schwartz}}\ and\ \bibinfo {author} {\bibfnamefont {D.}~\bibnamefont
  {Giulini}},\ }\bibfield  {title} {\bibinfo {title} {Post-{Newtonian}
  {Hamiltonian} description of an atom in a weak gravitational field},\ }\href
  {https://doi.org/10.1103/PhysRevA.100.052116} {\bibfield  {journal} {\bibinfo
   {journal} {Phys. Rev. A}\ }\textbf {\bibinfo {volume} {100}},\ \bibinfo
  {pages} {052116} (\bibinfo {year} {2019})}\BibitemShut {NoStop}%
\bibitem [{\citenamefont {Yudin}\ and\ \citenamefont
  {Taichenachev}(2018)}]{Yudin2018}%
  \BibitemOpen
  \bibfield  {author} {\bibinfo {author} {\bibfnamefont {V.}~\bibnamefont
  {Yudin}}\ and\ \bibinfo {author} {\bibfnamefont {A.}~\bibnamefont
  {Taichenachev}},\ }\bibfield  {title} {\bibinfo {title} {Mass defect effects
  in atomic clocks},\ }\href {https://doi.org/10.1088/1612-202X/aa9aa5}
  {\bibfield  {journal} {\bibinfo  {journal} {Laser Phys. Lett.}\ }\textbf
  {\bibinfo {volume} {15}},\ \bibinfo {pages} {035703} (\bibinfo {year}
  {2018})}\BibitemShut {NoStop}%
\bibitem [{\citenamefont {Mart\'{\i}nez-Lahuerta}\ \emph
  {et~al.}(2022)\citenamefont {Mart\'{\i}nez-Lahuerta}, \citenamefont {Eilers},
  \citenamefont {Mehlst\"aubler}, \citenamefont {Schmidt},\ and\ \citenamefont
  {Hammerer}}]{Martinez2022}%
  \BibitemOpen
  \bibfield  {author} {\bibinfo {author} {\bibfnamefont {V.~J.}\ \bibnamefont
  {Mart\'{\i}nez-Lahuerta}}, \bibinfo {author} {\bibfnamefont {S.}~\bibnamefont
  {Eilers}}, \bibinfo {author} {\bibfnamefont {T.~E.}\ \bibnamefont
  {Mehlst\"aubler}}, \bibinfo {author} {\bibfnamefont {P.~O.}\ \bibnamefont
  {Schmidt}},\ and\ \bibinfo {author} {\bibfnamefont {K.}~\bibnamefont
  {Hammerer}},\ }\bibfield  {title} {\bibinfo {title} {Ab initio quantum theory
  of mass defect and time dilation in trapped-ion optical clocks},\ }\href
  {https://doi.org/10.1103/PhysRevA.106.032803} {\bibfield  {journal} {\bibinfo
   {journal} {Phys. Rev. A}\ }\textbf {\bibinfo {volume} {106}},\ \bibinfo
  {pages} {032803} (\bibinfo {year} {2022})}\BibitemShut {NoStop}%
\bibitem [{\citenamefont {M\"untinga}\ \emph {et~al.}(2013)\citenamefont
  {M\"untinga}, \citenamefont {Ahlers}, \citenamefont {Krutzik}, \citenamefont
  {Wenzlawski}, \citenamefont {Arnold}, \citenamefont {Becker}, \citenamefont
  {Bongs}, \citenamefont {Dittus}, \citenamefont {Duncker}, \citenamefont
  {Gaaloul} \emph {et~al.}}]{Muentinga2013}%
  \BibitemOpen
  \bibfield  {author} {\bibinfo {author} {\bibfnamefont {H.}~\bibnamefont
  {M\"untinga}}, \bibinfo {author} {\bibfnamefont {H.}~\bibnamefont {Ahlers}},
  \bibinfo {author} {\bibfnamefont {M.}~\bibnamefont {Krutzik}}, \bibinfo
  {author} {\bibfnamefont {A.}~\bibnamefont {Wenzlawski}}, \bibinfo {author}
  {\bibfnamefont {S.}~\bibnamefont {Arnold}}, \bibinfo {author} {\bibfnamefont
  {D.}~\bibnamefont {Becker}}, \bibinfo {author} {\bibfnamefont
  {K.}~\bibnamefont {Bongs}}, \bibinfo {author} {\bibfnamefont
  {H.}~\bibnamefont {Dittus}}, \bibinfo {author} {\bibfnamefont
  {H.}~\bibnamefont {Duncker}}, \bibinfo {author} {\bibfnamefont
  {N.}~\bibnamefont {Gaaloul}}, \emph {et~al.},\ }\bibfield  {title} {\bibinfo
  {title} {Interferometry with {Bose-Einstein} condensates in microgravity},\
  }\href {https://doi.org/10.1103/PhysRevLett.110.093602} {\bibfield  {journal}
  {\bibinfo  {journal} {Phys. Rev. Lett.}\ }\textbf {\bibinfo {volume} {110}},\
  \bibinfo {pages} {093602} (\bibinfo {year} {2013})}\BibitemShut {NoStop}%
\bibitem [{\citenamefont {Dickerson}\ \emph {et~al.}(2013)\citenamefont
  {Dickerson}, \citenamefont {Hogan}, \citenamefont {Sugarbaker}, \citenamefont
  {Johnson},\ and\ \citenamefont {Kasevich}}]{Dickerson2013}%
  \BibitemOpen
  \bibfield  {author} {\bibinfo {author} {\bibfnamefont {S.~M.}\ \bibnamefont
  {Dickerson}}, \bibinfo {author} {\bibfnamefont {J.~M.}\ \bibnamefont
  {Hogan}}, \bibinfo {author} {\bibfnamefont {A.}~\bibnamefont {Sugarbaker}},
  \bibinfo {author} {\bibfnamefont {D.~M.~S.}\ \bibnamefont {Johnson}},\ and\
  \bibinfo {author} {\bibfnamefont {M.~A.}\ \bibnamefont {Kasevich}},\
  }\bibfield  {title} {\bibinfo {title} {Multiaxis inertial sensing with
  long-time point source atom interferometry},\ }\href
  {https://doi.org/10.1103/PhysRevLett.111.083001} {\bibfield  {journal}
  {\bibinfo  {journal} {Phys. Rev. Lett.}\ }\textbf {\bibinfo {volume} {111}},\
  \bibinfo {pages} {083001} (\bibinfo {year} {2013})}\BibitemShut {NoStop}%
\bibitem [{\citenamefont {Lan}\ \emph {et~al.}(2012)\citenamefont {Lan},
  \citenamefont {Kuan}, \citenamefont {Estey}, \citenamefont {Haslinger},\ and\
  \citenamefont {M\"uller}}]{Lan2012}%
  \BibitemOpen
  \bibfield  {author} {\bibinfo {author} {\bibfnamefont {S.-Y.}\ \bibnamefont
  {Lan}}, \bibinfo {author} {\bibfnamefont {P.-C.}\ \bibnamefont {Kuan}},
  \bibinfo {author} {\bibfnamefont {B.}~\bibnamefont {Estey}}, \bibinfo
  {author} {\bibfnamefont {P.}~\bibnamefont {Haslinger}},\ and\ \bibinfo
  {author} {\bibfnamefont {H.}~\bibnamefont {M\"uller}},\ }\bibfield  {title}
  {\bibinfo {title} {Influence of the coriolis force in atom interferometry},\
  }\href {https://doi.org/10.1103/PhysRevLett.108.090402} {\bibfield  {journal}
  {\bibinfo  {journal} {Phys. Rev. Lett.}\ }\textbf {\bibinfo {volume} {108}},\
  \bibinfo {pages} {090402} (\bibinfo {year} {2012})}\BibitemShut {NoStop}%
\bibitem [{\citenamefont {Roura}(2017)}]{Roura2017}%
  \BibitemOpen
  \bibfield  {author} {\bibinfo {author} {\bibfnamefont {A.}~\bibnamefont
  {Roura}},\ }\bibfield  {title} {\bibinfo {title} {Circumventing
  {Heisenberg}'s uncertainty principle in atom interferometry tests of the
  equivalence principle},\ }\href
  {https://doi.org/10.1103/PhysRevLett.118.160401} {\bibfield  {journal}
  {\bibinfo  {journal} {Phys. Rev. Lett.}\ }\textbf {\bibinfo {volume} {118}},\
  \bibinfo {pages} {160401} (\bibinfo {year} {2017})}\BibitemShut {NoStop}%
\bibitem [{\citenamefont {Overstreet}\ \emph {et~al.}(2018)\citenamefont
  {Overstreet}, \citenamefont {Asenbaum}, \citenamefont {Kovachy},
  \citenamefont {Notermans}, \citenamefont {Hogan},\ and\ \citenamefont
  {Kasevich}}]{Overstreet2018}%
  \BibitemOpen
  \bibfield  {author} {\bibinfo {author} {\bibfnamefont {C.}~\bibnamefont
  {Overstreet}}, \bibinfo {author} {\bibfnamefont {P.}~\bibnamefont
  {Asenbaum}}, \bibinfo {author} {\bibfnamefont {T.}~\bibnamefont {Kovachy}},
  \bibinfo {author} {\bibfnamefont {R.}~\bibnamefont {Notermans}}, \bibinfo
  {author} {\bibfnamefont {J.~M.}\ \bibnamefont {Hogan}},\ and\ \bibinfo
  {author} {\bibfnamefont {M.~A.}\ \bibnamefont {Kasevich}},\ }\bibfield
  {title} {\bibinfo {title} {Effective inertial frame in an atom
  interferometric test of the equivalence principle},\ }\href
  {https://doi.org/10.1103/PhysRevLett.120.183604} {\bibfield  {journal}
  {\bibinfo  {journal} {Phys. Rev. Lett.}\ }\textbf {\bibinfo {volume} {120}},\
  \bibinfo {pages} {183604} (\bibinfo {year} {2018})}\BibitemShut {NoStop}%
\bibitem [{\citenamefont {D'Amico}\ \emph {et~al.}(2017)\citenamefont
  {D'Amico}, \citenamefont {Rosi}, \citenamefont {Zhan}, \citenamefont
  {Cacciapuoti}, \citenamefont {Fattori},\ and\ \citenamefont
  {Tino}}]{DAmico2017}%
  \BibitemOpen
  \bibfield  {author} {\bibinfo {author} {\bibfnamefont {G.}~\bibnamefont
  {D'Amico}}, \bibinfo {author} {\bibfnamefont {G.}~\bibnamefont {Rosi}},
  \bibinfo {author} {\bibfnamefont {S.}~\bibnamefont {Zhan}}, \bibinfo {author}
  {\bibfnamefont {L.}~\bibnamefont {Cacciapuoti}}, \bibinfo {author}
  {\bibfnamefont {M.}~\bibnamefont {Fattori}},\ and\ \bibinfo {author}
  {\bibfnamefont {G.~M.}\ \bibnamefont {Tino}},\ }\bibfield  {title} {\bibinfo
  {title} {Canceling the gravity gradient phase shift in atom interferometry},\
  }\href {https://doi.org/10.1103/PhysRevLett.119.253201} {\bibfield  {journal}
  {\bibinfo  {journal} {Phys. Rev. Lett.}\ }\textbf {\bibinfo {volume} {119}},\
  \bibinfo {pages} {253201} (\bibinfo {year} {2017})}\BibitemShut {NoStop}%
\bibitem [{\citenamefont {Ufrecht}(2021)}]{Ufrecht2021GG}%
  \BibitemOpen
  \bibfield  {author} {\bibinfo {author} {\bibfnamefont {C.}~\bibnamefont
  {Ufrecht}},\ }\bibfield  {title} {\bibinfo {title} {Generalized
  gravity-gradient mitigation scheme},\ }\href
  {https://doi.org/10.1103/PhysRevA.103.023305} {\bibfield  {journal} {\bibinfo
   {journal} {Phys. Rev. A}\ }\textbf {\bibinfo {volume} {103}},\ \bibinfo
  {pages} {023305} (\bibinfo {year} {2021})}\BibitemShut {NoStop}%
\bibitem [{\citenamefont {Di~Pumpo}\ \emph {et~al.}(2022)\citenamefont
  {Di~Pumpo}, \citenamefont {Friedrich}, \citenamefont {Geyer}, \citenamefont
  {Ufrecht},\ and\ \citenamefont {Giese}}]{DiPumpo2022}%
  \BibitemOpen
  \bibfield  {author} {\bibinfo {author} {\bibfnamefont {F.}~\bibnamefont
  {Di~Pumpo}}, \bibinfo {author} {\bibfnamefont {A.}~\bibnamefont {Friedrich}},
  \bibinfo {author} {\bibfnamefont {A.}~\bibnamefont {Geyer}}, \bibinfo
  {author} {\bibfnamefont {C.}~\bibnamefont {Ufrecht}},\ and\ \bibinfo {author}
  {\bibfnamefont {E.}~\bibnamefont {Giese}},\ }\bibfield  {title} {\bibinfo
  {title} {Light propagation and atom interferometry in gravity and dilaton
  fields},\ }\href {https://doi.org/10.1103/PhysRevD.105.084065} {\bibfield
  {journal} {\bibinfo  {journal} {Phys. Rev. D}\ }\textbf {\bibinfo {volume}
  {105}},\ \bibinfo {pages} {084065} (\bibinfo {year} {2022})}\BibitemShut
  {NoStop}%
\bibitem [{\citenamefont {Alden}\ \emph {et~al.}(2014)\citenamefont {Alden},
  \citenamefont {Moore},\ and\ \citenamefont {Leanhardt}}]{Alden2014}%
  \BibitemOpen
  \bibfield  {author} {\bibinfo {author} {\bibfnamefont {E.~A.}\ \bibnamefont
  {Alden}}, \bibinfo {author} {\bibfnamefont {K.~R.}\ \bibnamefont {Moore}},\
  and\ \bibinfo {author} {\bibfnamefont {A.~E.}\ \bibnamefont {Leanhardt}},\
  }\bibfield  {title} {\bibinfo {title} {Two-photon ${{E1}}$-${{M1}}$ optical
  clock},\ }\href {https://doi.org/10.1103/PhysRevA.90.012523} {\bibfield
  {journal} {\bibinfo  {journal} {Phys. Rev. A}\ }\textbf {\bibinfo {volume}
  {90}},\ \bibinfo {pages} {012523} (\bibinfo {year} {2014})}\BibitemShut
  {NoStop}%
\bibitem [{\citenamefont {Janson}\ \emph {et~al.}(2024)\citenamefont {Janson},
  \citenamefont {Friedrich},\ and\ \citenamefont {Lopp}}]{Janson2024}%
  \BibitemOpen
  \bibfield  {author} {\bibinfo {author} {\bibfnamefont {G.}~\bibnamefont
  {Janson}}, \bibinfo {author} {\bibfnamefont {A.}~\bibnamefont {Friedrich}},\
  and\ \bibinfo {author} {\bibfnamefont {R.}~\bibnamefont {Lopp}},\ }\bibfield
  {title} {\bibinfo {title} {Finite pulse-time effects in long-baseline quantum
  clock interferometry},\ }\href {https://doi.org/10.1116/5.0178230} {\bibfield
   {journal} {\bibinfo  {journal} {AVS Quantum Sci.}\ }\textbf {\bibinfo
  {volume} {6}},\ \bibinfo {pages} {024403} (\bibinfo {year}
  {2024})}\BibitemShut {NoStop}%
\bibitem [{\citenamefont {Dolan}(2018)}]{Dolan2018}%
  \BibitemOpen
  \bibfield  {author} {\bibinfo {author} {\bibfnamefont {S.~R.}\ \bibnamefont
  {Dolan}},\ }\bibfield  {title} {\bibinfo {title} {Geometrical optics for
  scalar, electromagnetic and gravitational waves on curved spacetime},\ }\href
  {https://doi.org/10.1142/S0218271818430101} {\bibfield  {journal} {\bibinfo
  {journal} {Int. J. Mod. Phys. D}\ }\textbf {\bibinfo {volume} {27}},\
  \bibinfo {pages} {1843010} (\bibinfo {year} {2018})}\BibitemShut {NoStop}%
\bibitem [{\citenamefont {Wang}\ \emph {et~al.}(2021)\citenamefont {Wang},
  \citenamefont {Lu}, \citenamefont {Qin}, \citenamefont {Tan},\ and\
  \citenamefont {Shao}}]{Wang2021}%
  \BibitemOpen
  \bibfield  {author} {\bibinfo {author} {\bibfnamefont {Y.-J.}\ \bibnamefont
  {Wang}}, \bibinfo {author} {\bibfnamefont {X.-Y.}\ \bibnamefont {Lu}},
  \bibinfo {author} {\bibfnamefont {C.-G.}\ \bibnamefont {Qin}}, \bibinfo
  {author} {\bibfnamefont {Y.-J.}\ \bibnamefont {Tan}},\ and\ \bibinfo {author}
  {\bibfnamefont {C.-G.}\ \bibnamefont {Shao}},\ }\bibfield  {title} {\bibinfo
  {title} {Modeling gravitational wave detection with atom interferometry},\
  }\href {https://doi.org/10.1088/1361-6382/ac0236} {\bibfield  {journal}
  {\bibinfo  {journal} {Class. Quantum Grav.}\ }\textbf {\bibinfo {volume}
  {38}},\ \bibinfo {pages} {145025} (\bibinfo {year} {2021})}\BibitemShut
  {NoStop}%
\bibitem [{\citenamefont {Oko\l{}\'{o}w}(2020)}]{Okolow2020}%
  \BibitemOpen
  \bibfield  {author} {\bibinfo {author} {\bibfnamefont {A.}~\bibnamefont
  {Oko\l{}\'{o}w}},\ }\bibfield  {title} {\bibinfo {title} {{Does time always
  slow down as gravity increases?}},\ }\href
  {https://doi.org/10.1088/1361-6404/ab60bb} {\bibfield  {journal} {\bibinfo
  {journal} {Eur. J. Phys.}\ }\textbf {\bibinfo {volume} {41}},\ \bibinfo
  {pages} {023001} (\bibinfo {year} {2020})}\BibitemShut {NoStop}%
\bibitem [{\citenamefont {Giese}(2015)}]{Giese2015}%
  \BibitemOpen
  \bibfield  {author} {\bibinfo {author} {\bibfnamefont {E.}~\bibnamefont
  {Giese}},\ }\bibfield  {title} {\bibinfo {title} {Mechanisms of matter-wave
  diffraction and their application to interferometers},\ }\href
  {https://doi.org/10.1002/prop.201500020} {\bibfield  {journal} {\bibinfo
  {journal} {Fortschr. Phys.}\ }\textbf {\bibinfo {volume} {63}},\ \bibinfo
  {pages} {337} (\bibinfo {year} {2015})}\BibitemShut {NoStop}%
\bibitem [{\citenamefont {Schleich}\ \emph {et~al.}(2013)\citenamefont
  {Schleich}, \citenamefont {Greenberger},\ and\ \citenamefont
  {Rasel}}]{Schleich2013}%
  \BibitemOpen
  \bibfield  {author} {\bibinfo {author} {\bibfnamefont {W.~P.}\ \bibnamefont
  {Schleich}}, \bibinfo {author} {\bibfnamefont {D.~M.}\ \bibnamefont
  {Greenberger}},\ and\ \bibinfo {author} {\bibfnamefont {E.~M.}\ \bibnamefont
  {Rasel}},\ }\bibfield  {title} {\bibinfo {title} {{A representation-free
  description of the Kasevich{\textendash}Chu interferometer: a resolution of
  the redshift controversy}},\ }\href
  {https://doi.org/10.1088/1367-2630/15/1/013007} {\bibfield  {journal}
  {\bibinfo  {journal} {New J. Phys.}\ }\textbf {\bibinfo {volume} {15}},\
  \bibinfo {pages} {013007} (\bibinfo {year} {2013})}\BibitemShut {NoStop}%
\bibitem [{\citenamefont {Ufrecht}\ and\ \citenamefont
  {Giese}(2020)}]{Ufrecht2020}%
  \BibitemOpen
  \bibfield  {author} {\bibinfo {author} {\bibfnamefont {C.}~\bibnamefont
  {Ufrecht}}\ and\ \bibinfo {author} {\bibfnamefont {E.}~\bibnamefont
  {Giese}},\ }\bibfield  {title} {\bibinfo {title} {Perturbative operator
  approach to high-precision light-pulse atom interferometry},\ }\href
  {https://doi.org/10.1103/PhysRevA.101.053615} {\bibfield  {journal} {\bibinfo
   {journal} {Phys. Rev. A}\ }\textbf {\bibinfo {volume} {101}},\ \bibinfo
  {pages} {053615} (\bibinfo {year} {2020})}\BibitemShut {NoStop}%
\bibitem [{\citenamefont {Kleinert}\ \emph {et~al.}(2015)\citenamefont
  {Kleinert}, \citenamefont {Kajari}, \citenamefont {Roura},\ and\
  \citenamefont {Schleich}}]{Kleinert2015}%
  \BibitemOpen
  \bibfield  {author} {\bibinfo {author} {\bibfnamefont {S.}~\bibnamefont
  {Kleinert}}, \bibinfo {author} {\bibfnamefont {E.}~\bibnamefont {Kajari}},
  \bibinfo {author} {\bibfnamefont {A.}~\bibnamefont {Roura}},\ and\ \bibinfo
  {author} {\bibfnamefont {W.~P.}\ \bibnamefont {Schleich}},\ }\bibfield
  {title} {\bibinfo {title} {Representation-free description of light-pulse
  atom interferometry including non-inertial effects},\ }\href
  {https://doi.org/https://doi.org/10.1016/j.physrep.2015.09.004} {\bibfield
  {journal} {\bibinfo  {journal} {Phys. Rep.}\ }\textbf {\bibinfo {volume}
  {605}},\ \bibinfo {pages} {1} (\bibinfo {year} {2015})}\BibitemShut {NoStop}%
\bibitem [{\citenamefont {Giese}\ \emph {et~al.}(2014)\citenamefont {Giese},
  \citenamefont {Zeller}, \citenamefont {Kleinert}, \citenamefont {Meister},
  \citenamefont {Tamma}, \citenamefont {Roura},\ and\ \citenamefont
  {Schleich}}]{Giese2014}%
  \BibitemOpen
  \bibfield  {author} {\bibinfo {author} {\bibfnamefont {E.}~\bibnamefont
  {Giese}}, \bibinfo {author} {\bibfnamefont {W.}~\bibnamefont {Zeller}},
  \bibinfo {author} {\bibfnamefont {S.}~\bibnamefont {Kleinert}}, \bibinfo
  {author} {\bibfnamefont {M.}~\bibnamefont {Meister}}, \bibinfo {author}
  {\bibfnamefont {V.}~\bibnamefont {Tamma}}, \bibinfo {author} {\bibfnamefont
  {A.}~\bibnamefont {Roura}},\ and\ \bibinfo {author} {\bibfnamefont {W.~P.}\
  \bibnamefont {Schleich}},\ }\bibfield  {title} {\bibinfo {title} {The
  interface of gravity and quantum mechanics illuminated by {Wigner} phase
  space},\ }in\ \href {https://doi.org/10.3254/978-1-61499-448-0-171} {\emph
  {\bibinfo {booktitle} {Proceedings of the International Schoool of Physics
  ``Enrico Fermi,'' Course 188, Atom interferometry}}},\ \bibinfo {editor}
  {edited by\ \bibinfo {editor} {\bibfnamefont {G.~M.}\ \bibnamefont {Tino}}\
  and\ \bibinfo {editor} {\bibfnamefont {M.~A.}\ \bibnamefont {Kasevich}}}\
  (\bibinfo  {publisher} {IOS Press},\ \bibinfo {address} {Amsterdam},\
  \bibinfo {year} {2014})\ p.\ \bibinfo {pages} {171}\BibitemShut {NoStop}%
\bibitem [{\citenamefont {Roura}(2025)}]{Roura2025}%
  \BibitemOpen
  \bibfield  {author} {\bibinfo {author} {\bibfnamefont {A.}~\bibnamefont
  {Roura}},\ }\bibfield  {title} {\bibinfo {title} {Atom interferometer as a
  freely falling clock for time-dilation measurements},\ }\href
  {https://doi.org/10.1088/2058-9565/ad9e2e} {\bibfield  {journal} {\bibinfo
  {journal} {Quantum Sci. Technol.}\ }\textbf {\bibinfo {volume} {10}},\
  \bibinfo {pages} {025004} (\bibinfo {year} {2025})}\BibitemShut {NoStop}%
\bibitem [{\citenamefont {Kurasov}(1996)}]{Kurasov1996}%
  \BibitemOpen
  \bibfield  {author} {\bibinfo {author} {\bibfnamefont {P.}~\bibnamefont
  {Kurasov}},\ }\bibfield  {title} {\bibinfo {title} {{Distribution {Theory}
  for {Discontinuous} {Test} {Functions} and {Differential} {Operators} with
  {Generalized} {Coefficients}}},\ }\href
  {https://doi.org/10.1006/jmaa.1996.0256} {\bibfield  {journal} {\bibinfo
  {journal} {J. Math. Anal. Appl.}\ }\textbf {\bibinfo {volume} {201}},\
  \bibinfo {pages} {297} (\bibinfo {year} {1996})}\BibitemShut {NoStop}%
\bibitem [{\citenamefont {Le~Gou\"et}\ \emph {et~al.}()\citenamefont
  {Le~Gou\"et}, \citenamefont {Cheinet}, \citenamefont {Kim}, \citenamefont
  {Holleville}, \citenamefont {Clairon}, \citenamefont {Landragin},\ and\
  \citenamefont {Pereira Dos~Santos}}]{LeGouet2007}%
  \BibitemOpen
  \bibfield  {author} {\bibinfo {author} {\bibfnamefont {J.}~\bibnamefont
  {Le~Gou\"et}}, \bibinfo {author} {\bibfnamefont {P.}~\bibnamefont {Cheinet}},
  \bibinfo {author} {\bibfnamefont {J.}~\bibnamefont {Kim}}, \bibinfo {author}
  {\bibfnamefont {D.}~\bibnamefont {Holleville}}, \bibinfo {author}
  {\bibfnamefont {A.}~\bibnamefont {Clairon}}, \bibinfo {author} {\bibfnamefont
  {A.}~\bibnamefont {Landragin}},\ and\ \bibinfo {author} {\bibfnamefont
  {F.}~\bibnamefont {Pereira Dos~Santos}},\ }\bibfield  {title} {\bibinfo
  {title} {Influence of lasers propagation delay on the sensitivity of atom
  interferometers},\ }\href {https://doi.org/10.1140/epjd/e2007-00218-2}
  {\bibfield  {journal} {\bibinfo  {journal} {Eur. Phys. J. D}\ }\textbf
  {\bibinfo {volume} {44}},\ \bibinfo {pages} {419}}\BibitemShut {NoStop}%
\bibitem [{\citenamefont {B\"ohringer}\ \emph {et~al.}(2025)\citenamefont
  {B\"ohringer}, \citenamefont {Kienle},\ and\ \citenamefont
  {Lopp}}]{Boehringer2025}%
  \BibitemOpen
  \bibfield  {author} {\bibinfo {author} {\bibfnamefont {S.}~\bibnamefont
  {B\"ohringer}}, \bibinfo {author} {\bibfnamefont {F.}~\bibnamefont
  {Kienle}},\ and\ \bibinfo {author} {\bibfnamefont {R.}~\bibnamefont {Lopp}},\
  }\bibfield  {title} {\bibinfo {title} {Evolution of momentum-dependent
  observables under stochastic phase noise in {Rabi} oscillations},\ }\href
  {https://doi.org/10.1103/PhysRevResearch.7.023048} {\bibfield  {journal}
  {\bibinfo  {journal} {Phys. Rev. Res.}\ }\textbf {\bibinfo {volume} {7}},\
  \bibinfo {pages} {023048} (\bibinfo {year} {2025})}\BibitemShut {NoStop}%
\bibitem [{\citenamefont {M\"uller}\ \emph {et~al.}(2008)\citenamefont
  {M\"uller}, \citenamefont {Chiow},\ and\ \citenamefont {Chu}}]{Mueller2008}%
  \BibitemOpen
  \bibfield  {author} {\bibinfo {author} {\bibfnamefont {H.}~\bibnamefont
  {M\"uller}}, \bibinfo {author} {\bibfnamefont {S.-w.}\ \bibnamefont
  {Chiow}},\ and\ \bibinfo {author} {\bibfnamefont {S.}~\bibnamefont {Chu}},\
  }\bibfield  {title} {\bibinfo {title} {Atom-wave diffraction between the
  {Raman-Nath} and the {Bragg} regime: Effective rabi frequency, losses, and
  phase shifts},\ }\href {https://doi.org/10.1103/PhysRevA.77.023609}
  {\bibfield  {journal} {\bibinfo  {journal} {Phys. Rev. A}\ }\textbf {\bibinfo
  {volume} {77}},\ \bibinfo {pages} {023609} (\bibinfo {year}
  {2008})}\BibitemShut {NoStop}%
\bibitem [{\citenamefont {Clauser}(1988)}]{Clauser1988}%
  \BibitemOpen
  \bibfield  {author} {\bibinfo {author} {\bibfnamefont {J.~F.}\ \bibnamefont
  {Clauser}},\ }\bibfield  {title} {\bibinfo {title} {Ultra-high sensitivity
  accelerometers and gyroscopes using neutral atom matter-wave
  interferometry},\ }\href
  {https://doi.org/https://doi.org/10.1016/0378-4363(88)90176-3} {\bibfield
  {journal} {\bibinfo  {journal} {Physica B+C}\ }\textbf {\bibinfo {volume}
  {151}},\ \bibinfo {pages} {262} (\bibinfo {year} {1988})}\BibitemShut
  {NoStop}%
\bibitem [{\citenamefont {Giese}\ \emph {et~al.}(2013)\citenamefont {Giese},
  \citenamefont {Roura}, \citenamefont {Tackmann}, \citenamefont {Rasel},\ and\
  \citenamefont {Schleich}}]{Giese2013}%
  \BibitemOpen
  \bibfield  {author} {\bibinfo {author} {\bibfnamefont {E.}~\bibnamefont
  {Giese}}, \bibinfo {author} {\bibfnamefont {A.}~\bibnamefont {Roura}},
  \bibinfo {author} {\bibfnamefont {G.}~\bibnamefont {Tackmann}}, \bibinfo
  {author} {\bibfnamefont {E.~M.}\ \bibnamefont {Rasel}},\ and\ \bibinfo
  {author} {\bibfnamefont {W.~P.}\ \bibnamefont {Schleich}},\ }\bibfield
  {title} {\bibinfo {title} {Double {Bragg} diffraction: A tool for atom
  optics},\ }\href {https://doi.org/10.1103/PhysRevA.88.053608} {\bibfield
  {journal} {\bibinfo  {journal} {Phys. Rev. A}\ }\textbf {\bibinfo {volume}
  {88}},\ \bibinfo {pages} {053608} (\bibinfo {year} {2013})}\BibitemShut
  {NoStop}%
\bibitem [{\citenamefont {K\"uber}\ \emph {et~al.}(2016)\citenamefont
  {K\"uber}, \citenamefont {Schmaltz},\ and\ \citenamefont
  {Birkl}}]{Kueber2016}%
  \BibitemOpen
  \bibfield  {author} {\bibinfo {author} {\bibfnamefont {J.}~\bibnamefont
  {K\"uber}}, \bibinfo {author} {\bibfnamefont {F.}~\bibnamefont {Schmaltz}},\
  and\ \bibinfo {author} {\bibfnamefont {G.}~\bibnamefont {Birkl}},\ }\bibfield
   {title} {\bibinfo {title} {Experimental realization of double {Bragg}
  diffraction: robust beamsplitters, mirrors, and interferometers for
  {Bose-Einstein} condensates},\ }\href
  {https://doi.org/10.48550/arXiv.1603.08826} {\bibfield  {journal} {\bibinfo
  {journal} {arXiv:1603.08826}\ } (\bibinfo {year} {2016})}\BibitemShut
  {NoStop}%
\bibitem [{\citenamefont {Ahlers}\ \emph {et~al.}(2016)\citenamefont {Ahlers},
  \citenamefont {M\"untinga}, \citenamefont {Wenzlawski}, \citenamefont
  {Krutzik}, \citenamefont {Tackmann}, \citenamefont {Abend}, \citenamefont
  {Gaaloul}, \citenamefont {Giese}, \citenamefont {Roura}, \citenamefont {Kuhl}
  \emph {et~al.}}]{Ahlers2016}%
  \BibitemOpen
  \bibfield  {author} {\bibinfo {author} {\bibfnamefont {H.}~\bibnamefont
  {Ahlers}}, \bibinfo {author} {\bibfnamefont {H.}~\bibnamefont {M\"untinga}},
  \bibinfo {author} {\bibfnamefont {A.}~\bibnamefont {Wenzlawski}}, \bibinfo
  {author} {\bibfnamefont {M.}~\bibnamefont {Krutzik}}, \bibinfo {author}
  {\bibfnamefont {G.}~\bibnamefont {Tackmann}}, \bibinfo {author}
  {\bibfnamefont {S.}~\bibnamefont {Abend}}, \bibinfo {author} {\bibfnamefont
  {N.}~\bibnamefont {Gaaloul}}, \bibinfo {author} {\bibfnamefont
  {E.}~\bibnamefont {Giese}}, \bibinfo {author} {\bibfnamefont
  {A.}~\bibnamefont {Roura}}, \bibinfo {author} {\bibfnamefont
  {R.}~\bibnamefont {Kuhl}}, \emph {et~al.},\ }\bibfield  {title} {\bibinfo
  {title} {Double {Bragg} interferometry},\ }\href
  {https://doi.org/10.1103/PhysRevLett.116.173601} {\bibfield  {journal}
  {\bibinfo  {journal} {Phys. Rev. Lett.}\ }\textbf {\bibinfo {volume} {116}},\
  \bibinfo {pages} {173601} (\bibinfo {year} {2016})}\BibitemShut {NoStop}%
\bibitem [{\citenamefont {Schelfhout}\ \emph {et~al.}(2024)\citenamefont
  {Schelfhout}, \citenamefont {Hird}, \citenamefont {Hughes},\ and\
  \citenamefont {Foot}}]{Schelfhout2024}%
  \BibitemOpen
  \bibfield  {author} {\bibinfo {author} {\bibfnamefont {J.~S.}\ \bibnamefont
  {Schelfhout}}, \bibinfo {author} {\bibfnamefont {T.~M.}\ \bibnamefont
  {Hird}}, \bibinfo {author} {\bibfnamefont {K.~M.}\ \bibnamefont {Hughes}},\
  and\ \bibinfo {author} {\bibfnamefont {C.~J.}\ \bibnamefont {Foot}},\
  }\bibfield  {title} {\bibinfo {title} {Single-photon large-momentum-transfer
  atom interferometry scheme for {Sr} or {Yb} atoms with application to
  determining the fine-structure constant},\ }\href
  {https://doi.org/10.1103/PhysRevA.110.053309} {\bibfield  {journal} {\bibinfo
   {journal} {Phys. Rev. A}\ }\textbf {\bibinfo {volume} {110}},\ \bibinfo
  {pages} {053309} (\bibinfo {year} {2024})}\BibitemShut {NoStop}%
\end{thebibliography}%
\end{document}